\documentclass[journal=jacsat,manuscript=article]{achemso}

\usepackage[english]{babel}
\usepackage{amsfonts}
\usepackage{amsmath}
\usepackage{amsthm}
\usepackage{amssymb}
\usepackage{graphicx}
\usepackage{epspdfconversion}
\usepackage{subfigure}
\usepackage{cancel}
\usepackage{color}
\SectionNumbersOn

\newcommand{\rVec}{\mathbf{r}}
\newcommand{\rVecUnit}{\hat{\mathbf{r}}}

\author{Peter Poier}
\email{peter.poier@univie.ac.at}
\author{Christos N.~Likos}
\author{Richard Matthews}
\affiliation{Faculty of Physics, University of Vienna, Boltzmanngasse 5, A-1090 Vienna, Austria}

\title{Influence of Rigidity and Knot Complexity on the Knotting of Confined Polymers}

\begin{document}

\begin{abstract}
We employ computer simulations and thermodynamic integration to analyse the effects of bending rigidity and slit confinement on the free energy 
cost of tying knots, $\Delta F_{\rm knotting}$, on polymer chains under tension. A tension-dependent, non-zero optimal 
stiffness $\kappa_{\rm min}$ exists, for which $\Delta F_{\rm knotting}$ is minimal. 
For a polymer chain with several stiffness domains, each containing a large amount of monomers, the
domain with stiffness $\kappa_{\rm min}$ will be preferred by the knot. A {\it local} analysis of the bending in the interior of the knot reveals
that local stretching of chains at the braid region is responsible for the fact that the tension-dependent optimal stiffness has a non-zero
value. The reduction in $\Delta F_{\rm knotting}$ for a chain with optimal stiffness relative to the flexible chain can be enhanced by tuning the slit width of the 2D confinement and
increasing the knot complexity. The optimal stiffness itself is independent of the knot types
we considered, while confinement shifts it towards lower values.
\end{abstract}

\section{Introduction}
\label{sec:introduction}

Whilst in the macroscopic world it is clear that the effort needed to tie a knot in wire or string will always increase if it is made more rigid, for equivalent microscopic objects, polymers,
the same does not hold. Instead, it is found that the free energy cost of knotting a polymer, $\Delta F_{\rm knotting}$, has a minimum at a non-zero stiffness~\cite{Matthews2012}.
This finding is particularly interesting in the context of biological macromolecules, such as DNA or RNA, 
where on the one hand knotting is known to occur~\cite{Sogo1999,Arsuaga2002} 
and have significant effects on
key processes~\cite{Portugal1996,Deibler2007,Liu2009}, whilst on the other rigidity may depend sensitively on the base 
sequence~\cite{Hogan1983,Geggier2010,Johnson2013},
leading to varying flexibility along the polymer. Furthermore, there is  evidence of correlations between DNA stiffness and sites preferred by type 
II topoisomerases~\cite{Matthews2012}, enzymes that regulate knotting~\cite{Rybenkov1997}.

It is expected that the rigidity dependence of $\Delta F_{\rm knotting}$ will affect the behaviour 
of knots in DNA with non-uniform flexibility, for example by localising them in regions with
favourable stiffness. However,  previous work~\cite{Matthews2012} neglected a key qualitative feature of biological DNA, namely that it is typically highly 
confined~\cite{Emanuel2009,Jun2006,Purohit2005}. Confinement of a knotted polymer in a good solvent may significantly affect its properties. For example, in contrast to three
dimensions where they are weakly localised,  knots in polymers adsorbed on a surface are strongly localised~\cite{orlandini2009,ercolini:prl:2007}. Considering the properties of
polymers confined in a slit, simulations of DNA  found a non-monotonic dependence of the knotting probability on the slit width~\cite{Micheletti2012} and for flexible polymers
evidence was found that the particular topology is important~\cite{Matthews2011}. Whilst previous work on knotting in confinement has focussed on polymers that have one specific
stiffness, here we apply a simple model for a polymer chain under tension, to investigate the 
dependence of $\Delta F_{\rm knotting}$ on rigidity for various widths of the geometrical confinement. 
We find that a local stretching of the chains at the braiding region of the knots is responsible for the fact that the
optimal bending rigidity for knot formation, $\kappa_{\rm min}$, differs from zero. Geometric confinement, however, pushes this optimal rigidity
towards smaller values. The effect of confinement on $\kappa_{\rm min}$, as well as the amount 
by which  $\Delta F_{\rm knotting}$ is reduced for the optimal rigidity $\kappa_{\rm min}$ depends sensitively on the tension applied to the polymer chain.

The rest of the paper is organized as follows: We first present our model and details about the simulation in Section \ref{sec:model}. 
In Section \ref{sec:observables} we define and explain the observables that have been measured in our simulations. In particular, we define our 
notion of {\it bending} of the polymer chain and establish its connection to $\Delta F_{\rm knotting}$. Section \ref{sec:localBending} 
introduces the analysis of the {\it local bending} in the interior of the knot, which is carried out to investigate which part of the knot is responsible 
for the reduction of $\Delta F_{\rm knotting}$ for polymers with non-zero bending stiffness $\kappa_{\rm min}$ relative to a fully flexible chain. 
We present our results in Section \ref{sec:results}, whereas in Section \ref{sec:conclusions} we summarize and draw our conclusions.

\section{Model and simulation details}
\label{sec:model}
For the polymer chain (linear or knotted), we employ a standard,
self-avoiding bead-spring model with rigidity $\kappa$, confined parallel to the $(x,z)$-plane and under tension $\tau$. The interaction part of the
Hamiltonian thus reads as:
\begin{eqnarray}
\nonumber
V\left(\lbrace \rVec_i \rbrace \right) & = & \kappa \sum_i \left(1-\rVecUnit_{i-1,i} \cdot \rVecUnit_{i,i+1} \right)
- \frac{k R_0^2}{2}\sum_i \ln\left[1- \left(\frac{\rVec_{i,i+1}}{R_0}\right)^2\right]
\\
& + & 4 \epsilon\sum_{j>i}\sum_i \left[ \left( \frac{\sigma}{r_{i,j}}\right)^{12} - \left(\frac{\sigma}{r_{i,j}} \right)^6 + \frac{1}{4}\right]\Theta\left(2^{1/6}\sigma - r_{i,j}\right)
+ k_{\rm B}T\sum_i \left(\frac{y_i}{d} \right)^2 - \tau L_z .
\label{poten:eq}
\end{eqnarray}
In Eq.\ (\ref{poten:eq}), $\rVec_{i,j}=\rVec_j- \rVec_i$ is the vector from bead $i$ to bead $j$, located at position vectors $\rVec_i$ and $\rVec_j$, 
respectively, with the unit vector $\rVecUnit_{i,j}= {\rVec_{i,j}}/{|\rVec_{i,j}|}$. The first term represents the bending energy of the chain, where $\kappa$ is the bending rigidity. 
The second and third terms are the connectivity and steric terms, respectively, whereas $\Theta(\omega)$ is the Heaviside step function of $\omega$, 
which renders the Lennard-Jones potential purely repulsive. We choose $\epsilon = k_{\rm B} T$, $k = 30 k_{\rm B} T /\sigma^2$, and $R_0= 1.5 \sigma$, 
preventing the chain from crossing itself and thus conserving its topology. The chain is confined in a slit parallel to the ($x,z$) plane, which is realized 
via a harmonic external potential acting on the $y$-component
of the coordinate of each monomer, expressed by the fourth term in Eq.~(\ref{poten:eq}). The last term applies a 
tension $\tau$ on the chain along the $z$-direction of the setup, with $L_z$ denoting the extension of the chain along this direction.

We used the LAMMPS simulation package~\cite{Plimpton19951} to carry out constant-$N\tau T$ Molecular Dynamics (MD) simulations. 
The polymers consist of $N=256$ monomers for the chains simulated at tensions $\tau = 0.8\,k_{\rm B}T/\sigma$ and $\tau = 0.4\,k_{\rm B}T/\sigma$,
and of $N=512$ monomers for the simulations 
at tensions $\tau = 0.2\,k_{\rm B}T/\sigma$ and $\tau = 0.1\,k_{\rm B}T/\sigma$. 
The longer polymer chains at the two smaller tensions are necessary due to the larger knot size for these tension values. 
The chains are placed in a simulation box with volume $V=100 \sigma \times 150\sigma \times L_z$ at the two higher tensions and $V=200 \sigma \times 300\sigma \times L_z$ for the lower tensions. The tension $\tau$ is realized via a barostat coupled 
to the fluctuating $z$-length of the simulation box, whilst the box lengths in the $x$- and $y$-directions are fixed to $100\sigma$ and $150\sigma$ for the higher tensions $200\sigma$ and $300\sigma$ for the lower tensions respectively. The polymer is connected across the periodic
boundary conditions in the $z$-direction to guarantee that the knot is preserved. We also use periodic boundary conditions in the $x$-direction, while for the $y$-direction confinement prevents the polymer chain from getting outside the simulation-box. For both the thermostat and the barostat, we used a 
Nose-Hoover chain with 3 degrees of freedom~\cite{Tuckerman2001}. With $m$ denoting the monomer mass 
and $\beta = (k_{\rm B}T)^{-1}$, $t_0 = \sqrt{m\sigma^2\beta}$ sets the unit of time. We integrated the equations of motion with a 
timestep $\Delta t= 10^{-3}t_0$. The equilibration time was $2\times10^7$ timesteps, and data were collected during a total of $3\times10^8$ timesteps.

\section{Definition and interpretation of Observables}
\label{sec:observables}

{\it Definition and physical interpretation of $\Delta F_{\rm knotting}(\kappa)$}: Let $F_{\rm lin, knot}(\kappa)$ 
be the free energies of an unknotted and a knotted chain, respectively, for given $\kappa$, $\tau$ and chain length. 
We define $\Delta F_{\rm knotting} (\kappa) \equiv F_{\rm knot}(\kappa)-F_{\rm lin}(\kappa)$, 
a quantity that gives a measure for the effort to tie a knot into the chain of $N$ monomers. In our simulations, we do
not calculate the absolute value for $\Delta F_{\rm knotting} (\kappa)$ but its value relative to $\Delta F_{\rm knotting}(\kappa = 0)$ of a flexible chain,
a procedure that removes the $N$-dependence for a linear and a knotted chain of the same degree of polymerization $N$.
Thus, we calculate $\Psi(\kappa) \equiv \Delta F_{\rm knotting}(\kappa) -  \Delta F_{\rm knotting}(0)$, which has a direct physical interpretation: 
Let us consider a long polymer chain with various domains, which differ by their respective bending stiffness $\kappa_i$ and the number 
of monomers they contain $N_i$. Then, the quantity $\Psi(\kappa)$ allows us to predict the probability for the knot to be found in the $i$-th domain 
relative to the probability for it being in the $j$-th domain as:
\begin{eqnarray}
\frac{P_i}{P_j}= \frac{N_i}{N_j}\exp\left[-\beta \left(\Psi(\kappa_i)-\Psi(\kappa_j) \right)\right].
\label{ProbI:eq}
\end{eqnarray}

An implicit assumption entering Eq.~(\ref{ProbI:eq}) is that the length of the chain segments, $N_i$, is much longer than the knot size $N_K$, so that for most configurations 
the part of the polymer chain that is affected by the knot is localized in only one of the domains. Due to the applied tension $\tau$ the knotted part of the chain will always remain
finite. In Ref.\cite{Matthews2012}, it was shown that at fixed $\kappa$ the size of the knot does not scale with $N$, 
the number of monomers on the chain, but rather as $N_K\sim \left(k_B T/\tau\right)^\alpha$ with some exponent $\alpha$. 
The reason for this, is that the stretched polymer forms a series of tension blobs,\cite{rubinstein2003polymer} 
whose size scales with the tension $\tau$ but not with $N$. 
The knot can only be in one of those blobs, therefore $N_K$ will be finite for any non-zero tension $\tau$. 
Therefore, for sufficiently large $N_i$, Eq.~(\ref{ProbI:eq}) indeed provides a good prediction for $P_i$. In Ref. \cite{Matthews2012} the prediction 
of Eq.~(\ref{ProbI:eq}) was tested for a simulation of a polymer chain with two stiffness domains.

{\it Definition and interpretation of the chain's bending $\hat{B}$}:  With $\hat {\bf r}_{j,j+1}$ denoting the unit vector between monomers $j$ and $j+1$, 
we define the quantity 
\begin{eqnarray}
{\hat B}\equiv \sum_i \left(1-\rVecUnit_{i-1,i} \cdot \rVecUnit_{i,i+1} \right)
\label{bendingDef:eq}
\end{eqnarray}
for any configuration of its monomers and call it the \textit{bending} of the chain;
evidently it holds that $\hat B \geq 0$. One can check that this definition of the bending is sensible for various special cases. For instance, for a configuration with a straight polymer
chain this definition gives the minimum value for $\hat B$, namely $\hat B= 0$. The contribution to $V\left(\lbrace \rVec_i \rbrace \right)$ defined in (\ref{poten:eq}) due to the bending
stiffness is $\kappa \hat B$. Moreover $B_T(\kappa) \equiv \langle {\hat B}\rangle$ is the thermodynamic expectation value of the same 
for a chain of topology $T \in \{{\rm knot},{\rm linear}\}$.

With $F_{T}(\kappa)$ denoting the free energy of the polymer, it holds that $\partial F_T(\kappa)/\partial\kappa = B_T(\kappa)$. 
This allows us to calculate how the cost of knotting changes with bending stiffness 
$\kappa$: Introducing  $\Delta B(\kappa) \equiv B_{\rm knot}(\kappa) - B_{\rm linear}(\kappa)$, 
it follows that $\partial \Delta F_{\rm knotting}(\kappa)/\partial \kappa = \Delta B(\kappa)$.
The free energy cost of knotting is therefore: 
\begin{eqnarray}
\Delta F_{\rm knotting}(\kappa) =& \Delta F_{\rm knotting}(0)
+ \int_0^{\kappa}\Delta B(\kappa'){\rm d}\kappa',\nonumber
\end{eqnarray}
and thus
\begin{eqnarray}
\Psi(\kappa) =&  \int_0^{\kappa}\Delta B(\kappa'){\rm d}\kappa'.
\label{integrate:eq}
\end{eqnarray}

The existence of a minimum of $\Psi(\kappa)$ for $\kappa \ne 0$ will depend on the sign of the slope of $\Psi(\kappa)$
at $\kappa = 0$, $\Delta B(0)$. If $\Delta B(0) < 0$, we expect an optimal knotting rigidity $\kappa_{\rm min} \ne 0$, whereas we anticipate a
monotonically increasing function $\Psi(\kappa)$ in the opposite case, $\Delta B(0) \geq 0$.  
This is a reasonable expectation, since we will always obtain $\Delta B(\kappa)>0$ for sufficiently stiff chains. 
Indeed, for $\beta\kappa \gg 1$, a linear polymer will adopt an almost straight configuration  with $\hat{B}\approx 0$, as configurations 
with non-zero $\hat{B}$ are penalized by a high bending energy. $\hat{B}=0$ is unique for the straight configuration, which is of course not knotted. 
Therefore $\Delta B(\kappa)>0$ will hold for all knots in the $\beta\kappa \gg 1$ regime.

To illustrate this further, let us for example consider a stiff polymer chain with a trefoil knot. Its minimal energy is obtained for a straight polymer chain with 
an approximately circular domain at the location of the knot. The circle is tangent to the braid point and 
contains $N_{\rm K}(\kappa) \cong \sqrt{2\pi^2\kappa/(\tau\ell)}$ monomers,
where $\ell \cong \sigma$ is the bond length \cite{Gallotti2007}. Accordingly, $\Delta B(\kappa) \cong \sqrt{2\pi^2\tau\ell/\kappa} > 0$ in this limit.

\section{Analysis of the local bending in the knotted domain}
\label{sec:localBending}

Having in mind the 
goal of localizing which part of the polymer in the vicinity of the knot is contributing to an increased or decreased average bending and therefore 
to a non vanishing $\Delta B(\kappa)$, we need to determine the knotted domain on the polymer chain. We define the knotted state of open subsections
by introducing a topologically neutral closure scheme, which transforms an open string into a ring polymer that has a mathematically well defined topological 
state~\cite{Tubiana2011, Marcone2007, Marcone2005}. We used a scheme where the end points of the open polymer are connected to a sphere at infinity in the direction of the
vector from the centroid to the respective end point. The knotted part of the polymer chain is then the smallest domain 
for which the closure yields a ring polymer that has the correct Alexander polynomial.

Once the ends of the knot have been identified, we introduce a new enumeration scheme for the monomers, denoted by the Greek integer index $\alpha$, which
can have positive as well as negative values. The two monomers in the {\it interior} of the knot that lie $n$ bonds away from the endpoints 
obtain the index $\alpha = -n$, whereas the two monomers to be found $n$ bonds away at the {\it exterior} of the knot are assigned the index $\alpha = n$; accordingly,
$\alpha = 0$ for the two endpoints of the knot. There exist, thus, for every value of $\alpha$ two position vectors ${\bf r}_{\alpha}^j$, $j = L,R$,
where $L/R$ denotes whether the monomer is at separation $\alpha$ from the left/right endpoint of the knot. Accordingly, we define
the local bending contribution from the two monomers carrying the index $\alpha$ as:
\begin{eqnarray}
\hat b_{\alpha} \equiv \sum_{j = L,R}\left(1 - \rVecUnit_{\alpha-1,\alpha}^j \cdot \rVecUnit_{\alpha,\alpha+1}^j\right).
\label{locanbend:eq}
\end{eqnarray}
For $\alpha<0$, $\hat b_{\alpha}$ measures the {\it local} bending of angles {\it in the interior} of the knot, while for $\alpha>0$, $\hat b_{\alpha}$ 
measures the {\it local} bending {\it outside} of the domain that was identified as knotted. The domain limits for $\alpha \in [\alpha_{\rm min}, \alpha_{\rm max}]$ 
depend on the instantaneous configuration, as the number of monomers on the knot $N_K$ determines how negative $\alpha$ can become. 
The range of $\alpha$ is constant, $\alpha_{\rm max} - \alpha_{\rm min} +1  = N/2$.

We introduce a characteristic function $\hat\chi_{\alpha} = 1$ or $0$ depending
on whether the index $\alpha$ occurs for a given conformation or not and define the \textit{local} bending difference between a knotted and a linear chain as
\begin{eqnarray}
\Delta b_{\alpha}(\kappa) \equiv \left\langle \hat\chi_{\alpha}\left(\hat b_{\alpha} - b_{\rm linear}(\kappa)\right)\right\rangle.
\label{deltabalpha:eq}
\end{eqnarray}
In Eq. (\ref{deltabalpha:eq}) above, $b_{\rm linear}(\kappa) \equiv 2B_{\rm linear}(\kappa)/N$ is the thermodynamic average of the bending of two angles on a linear chain.
It follows that 
\begin{eqnarray}
\Delta B(\kappa) = \sum_{\alpha = -M}^M \Delta b_{\alpha}(\kappa),
\label{integral:eq}
\end{eqnarray}
where $M = N/2$ is the smallest value guaranteeing $[\alpha_{\rm min}, \alpha_{\rm max}] \subset [-M,M]$ for all polymer configurations.
\section{Results}
\label{sec:results}

\begin{figure}[htp]
\begin{center}
\subfigure
{
\includegraphics[width=7.5cm]{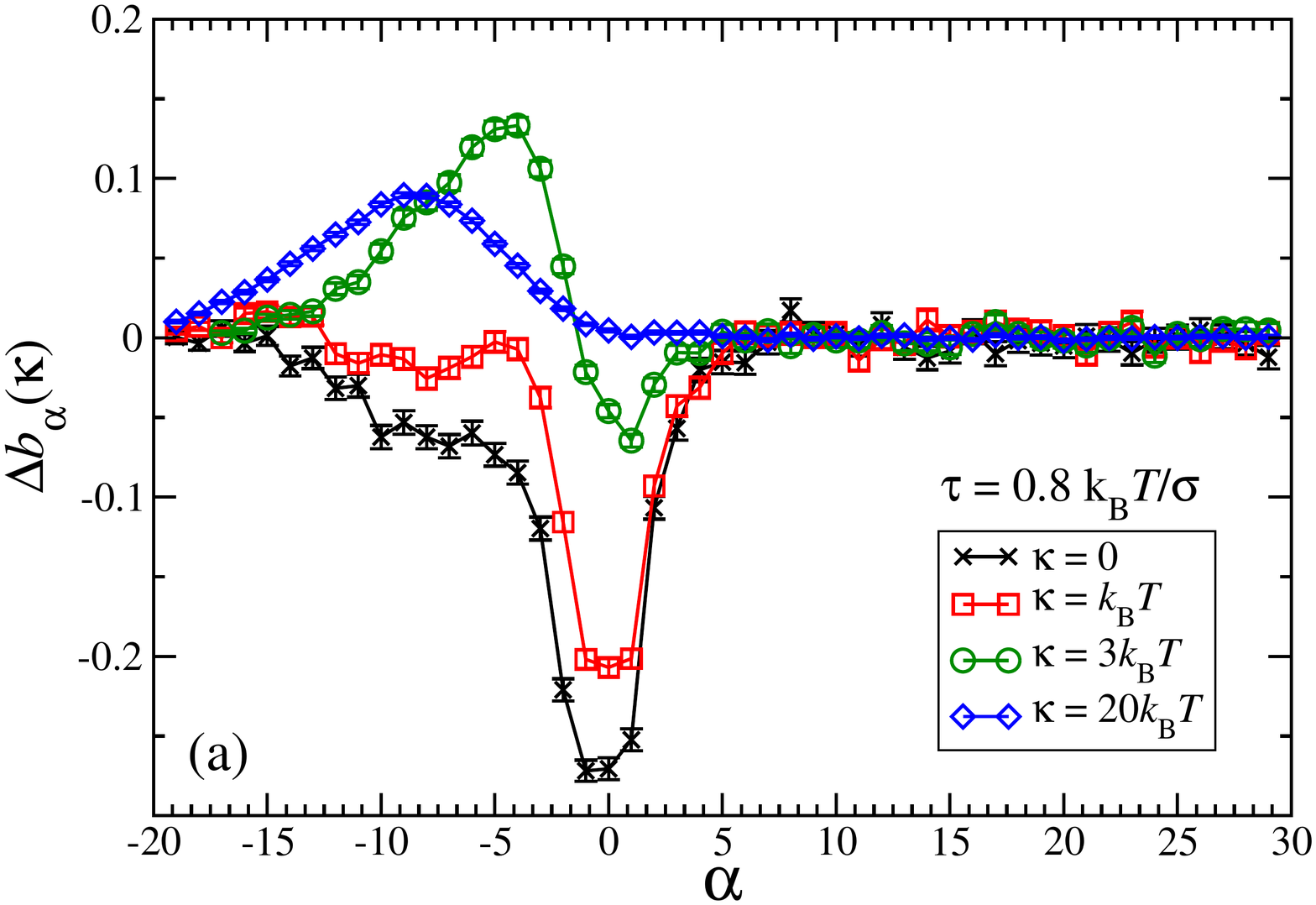} 
}
{
\includegraphics[width=7.5cm]{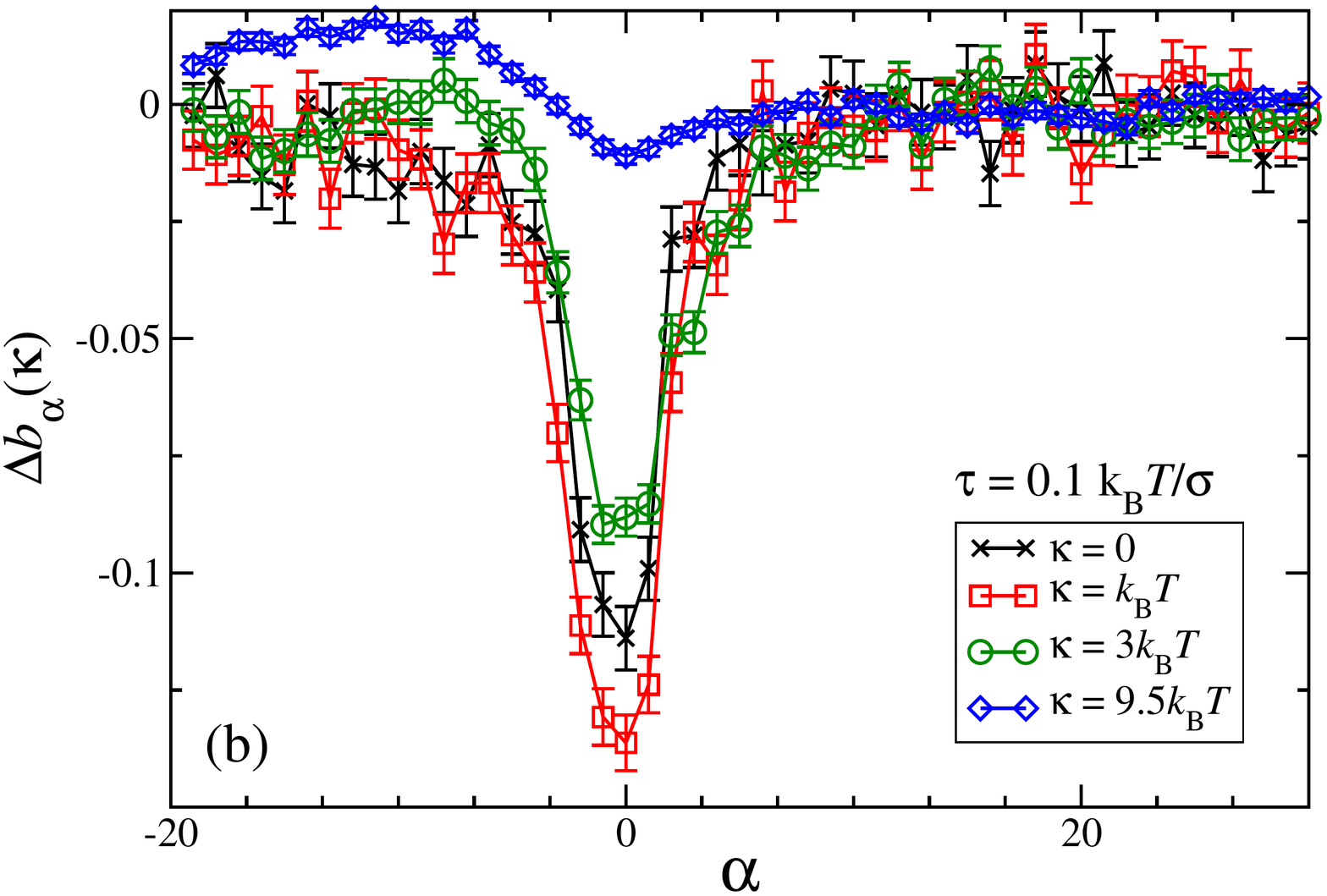} 
}
\vspace{0.3cm}
\subfigure
{
\includegraphics[width=7.5cm]{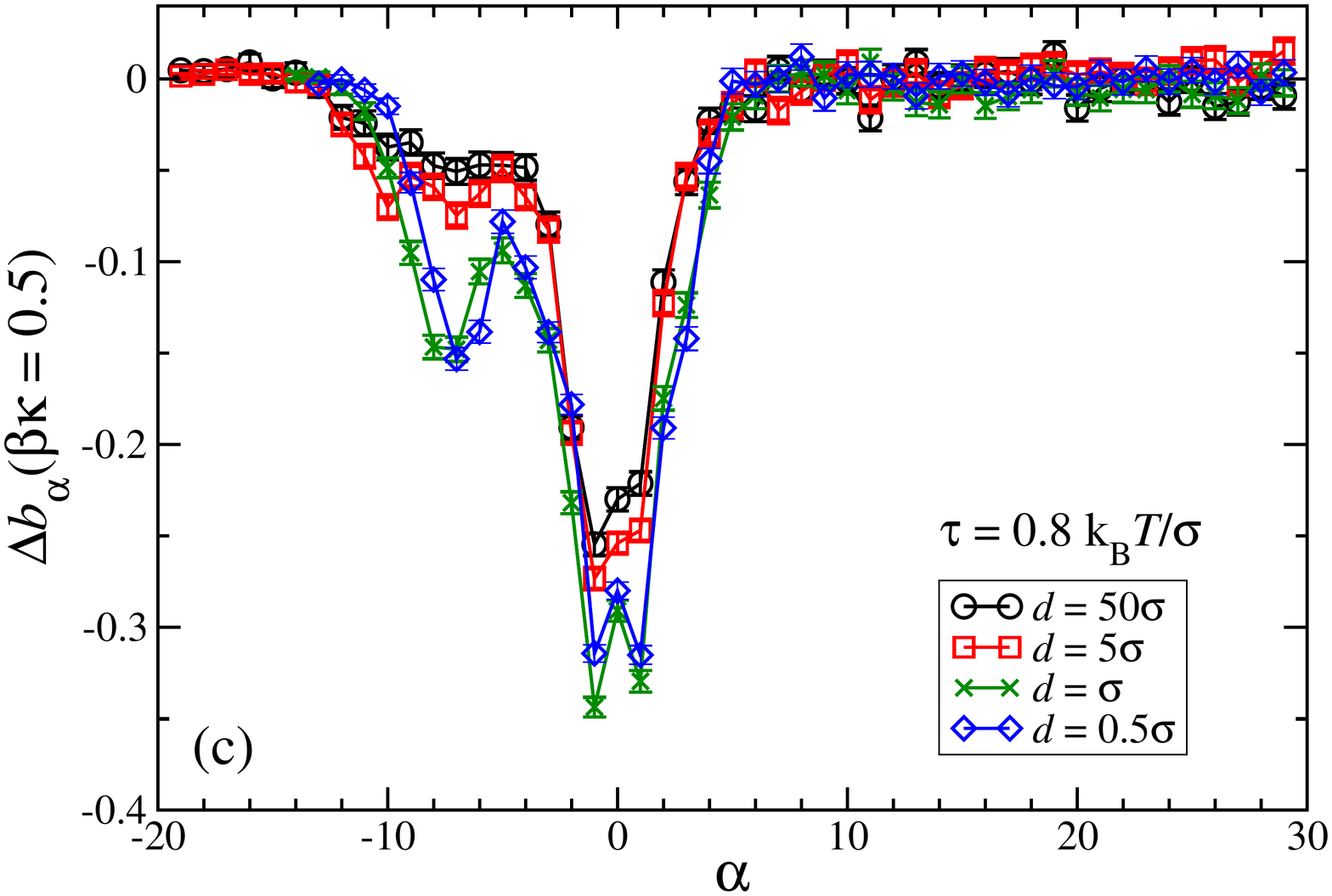} 
}
{
\includegraphics[width=7.5cm]{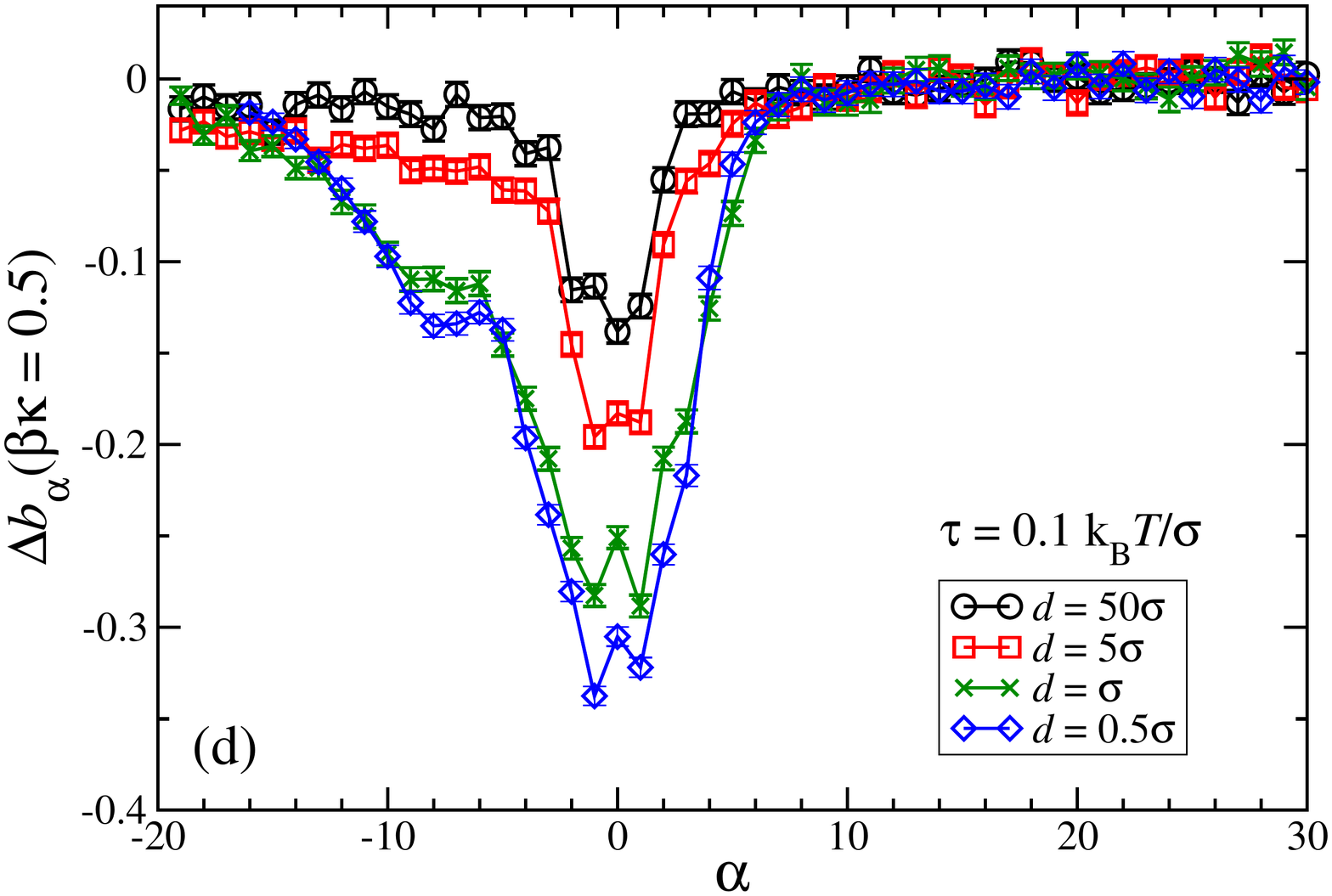} 
}
\end{center}
\caption{(a), (b): The average local bending difference $\Delta b_{\alpha}(\kappa)$ (see text) between knotted and linear unconfined polymers 
of different rigidities, as indicated in the legend. (c), (d): The same quantity at fixed bending rigidity $\kappa = 0.5\,k_{\rm B}T$ for different degrees 
of slit confinement, as indicated in the legend. In panels (a) and (c) data for the tension $\tau = 0.8\,k_{\rm B}T/\sigma$ are shown, 
while in panels (b) and (d) for $\tau = 0.1\,k_{\rm B}T/\sigma$.}
\label{AddBendingPerDeltaPlots}
\end{figure}

\begin{figure}[htp]
\includegraphics[width=6cm]{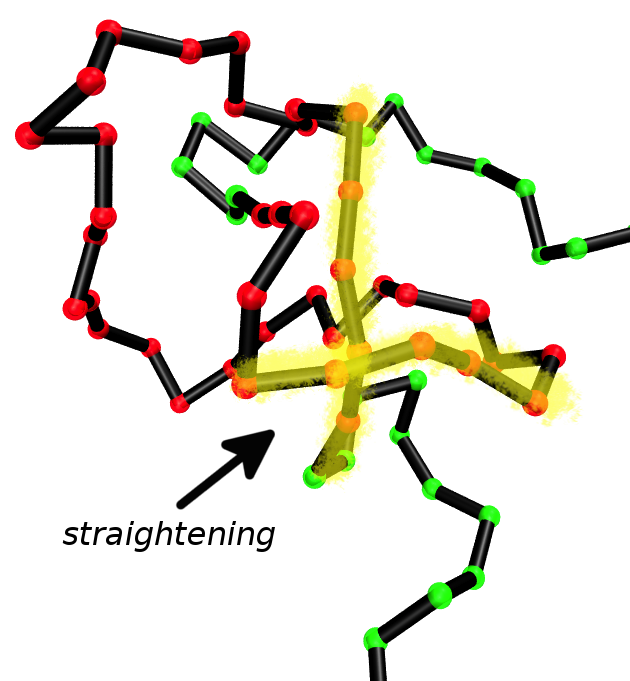} 
\caption{Simulation snapshot of the knotted domain of an unconfined, fully flexible polymer chain with a trefoil knot under the tension 
$\tau = 0.8\,k_{\rm B}T/\sigma$. 
The straightened-out segments in the vicinity of the strand crossings are highlighted.}
\label{supprPlot}
\end{figure}

Results for the local bending of unconfined polymers are shown in Figs.~\ref{AddBendingPerDeltaPlots}(a) 
and \ref{AddBendingPerDeltaPlots}(b) for two different values of the applied tension. As can be seen, the quantity
$\Delta b_{\alpha}(\kappa)$ vanishes within less than 10 beads outside the knot; outside this region,
a knotted chain hardly differs in its bending from an unknotted one. According to Eqs.~(\ref{integrate:eq}) and (\ref{integral:eq}),
the area under the curves in Figs.~\ref{AddBendingPerDeltaPlots}(a) and \ref{AddBendingPerDeltaPlots}(b)
determines the slope or $\Psi(\kappa)$ at any
$\kappa$-value. 
The major negative contribution to $\Delta B(\kappa)$ for $\kappa = 0$ comes from a small number of monomers close to the knot ends,
i.e., in the braiding-region of the knot, in which monomers start getting into the knotted domain.
The bending suppression is therefore related to the interaction of the different strands at the crossings, which effectively confines the strands 
and hence reduces their random bending.
Thus, the negative slope of $\Delta F_{\rm knotting}(\kappa)$  at $\kappa = 0$ arises from this {\it additional} 
straightening of the knotted chain with respect to its flexible, unknotted counterpart. As $\Delta B(\kappa=0)$, 
which is equal to the area under the respective curves in 
Figs.~\ref{AddBendingPerDeltaPlots}(a) and \ref{AddBendingPerDeltaPlots}(b), is negative, 
a flexible knotted chain is, on average, {\it less bent} than
its unknotted counterpart, contrary to the intuitive expectation that knotting inevitably {\it increases} the total bending of a polymer.
In Fig.\ \ref{supprPlot} we show a simulation snapshot of the knotted domain of a flexible chain, in which the parts of the molecule that
get `straightened out' due to the knot are highlighted. As $\kappa$ grows, we eventually reach the intuitively expected regime
in which knotting increases bending, see, e.g., the curve for $\kappa = 20\,k_{\rm B}T$ in Fig.\ \ref{AddBendingPerDeltaPlots}(a),
for which $\Delta B(\kappa) > 0$.

Comparing the curves of Fig.~\ref{AddBendingPerDeltaPlots}(a) and (b), one sees that $\Delta B(\kappa)$ for $\kappa = 0$ 
is more negative for $\tau = 0.8\,k_{\rm B}T/\sigma$ than for  $\tau = 0.1\,k_{\rm B}T/\sigma$. This is due to the fact that at smaller $\tau$ the 
braid region is looser and the bending suppression due to the different strands at the crossings is reduced. On the contrary, if we increase $\kappa$, 
we arrive at a regime where $\Delta B(\kappa)$ is more negative for smaller tensions $\tau$. At $\beta\kappa=3.0$, $\Delta B$ 
is already positive for $\tau = 0.8\,k_{\rm B}T/\sigma$, due to the positive contribution of $\Delta b_{\alpha}(\kappa)$ in the interior of the knot, 
which arises as the knot enforces the polymer to form a loop. The bending throughout a loop is larger if the loop is smaller. 
For $\tau = 0.1\,k_{\rm B}T/\sigma$ the knot size is increased, which is the reason why the positive $\Delta b_{\alpha}(\kappa)$ 
contribution in the interior of the knot is significantly smaller than for $\tau = 0.8\,k_{\rm B}T/\sigma$. Accordingly, for lower tensions the negative net result for $\partial \Psi(\kappa)/\partial\kappa$ persists for higher $\kappa$-values than for higher 
tensions. As can be seen in Fig.~\ref{AddBendingPerDeltaPlots}(b), we see a reversal of $\partial \Psi(\kappa)/\partial\kappa$ from negative to positive values only at a rigidity as high as $\beta\kappa \cong 9$.

\begin{figure}[htp]
\subfigure{
{
\includegraphics[width=7.5cm]{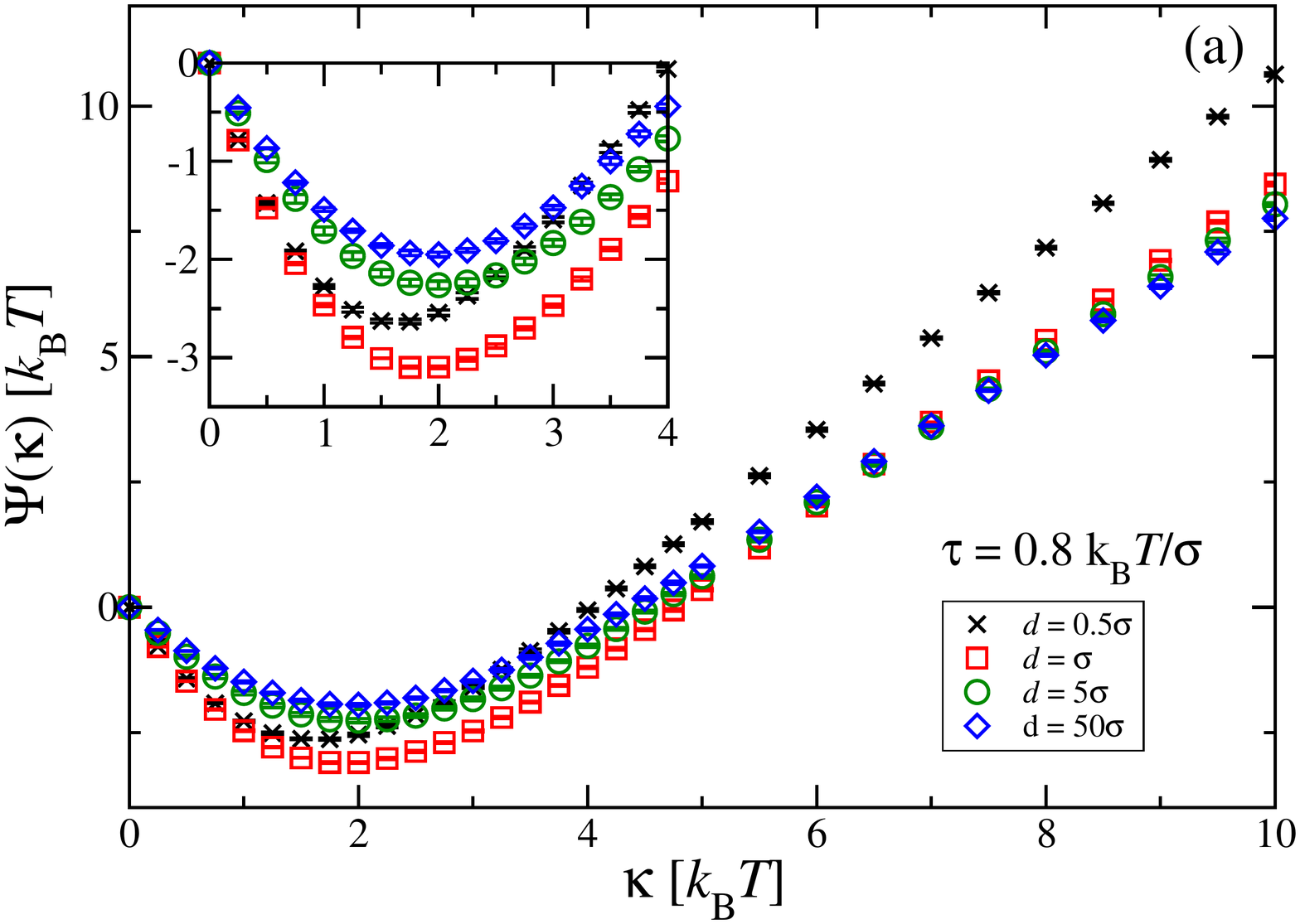} 
}
{
\includegraphics[width=7.5cm]{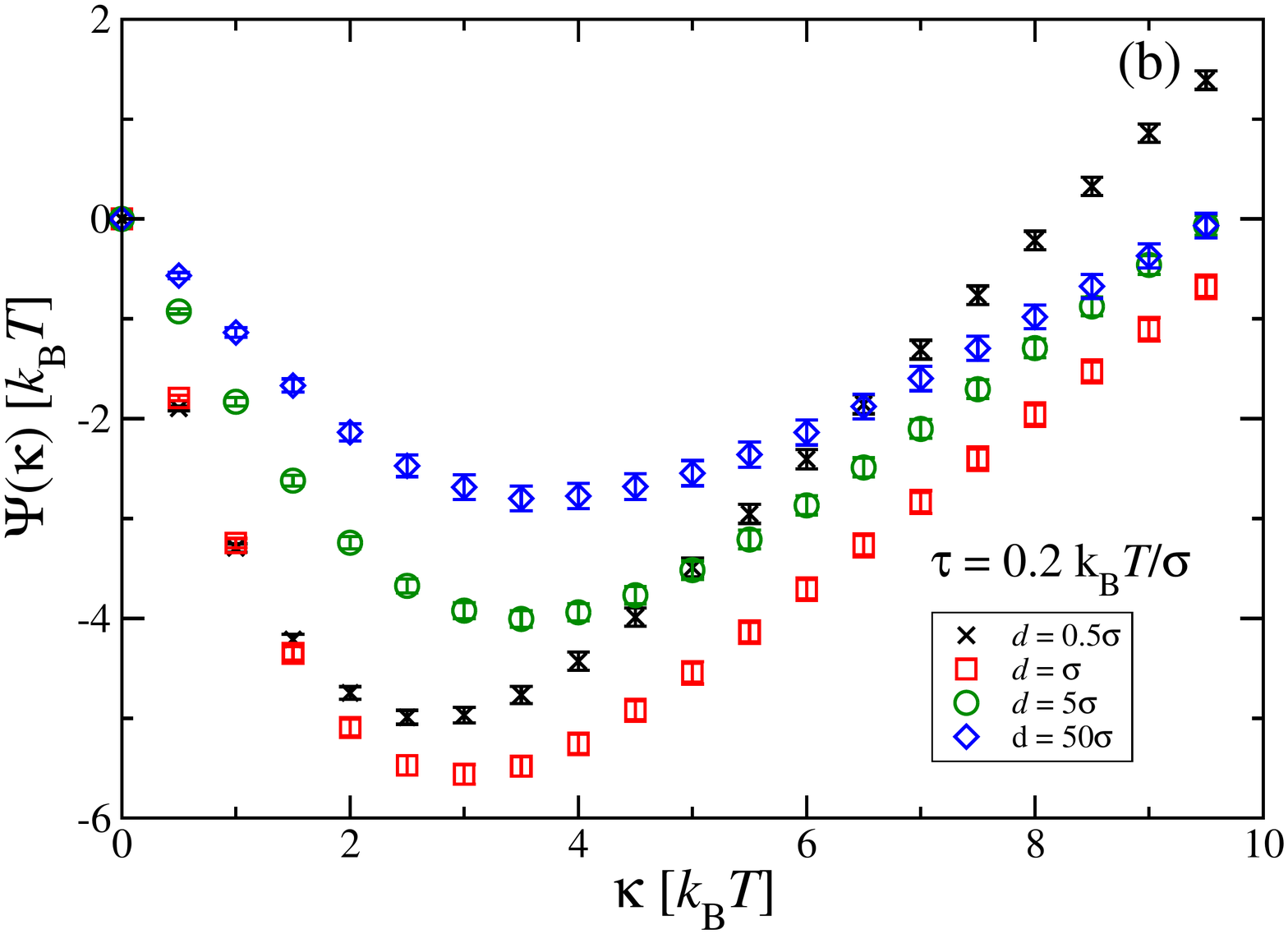} 
}
\vspace{0.3cm}
}
\subfigure{
{
\includegraphics[width=7.5cm]{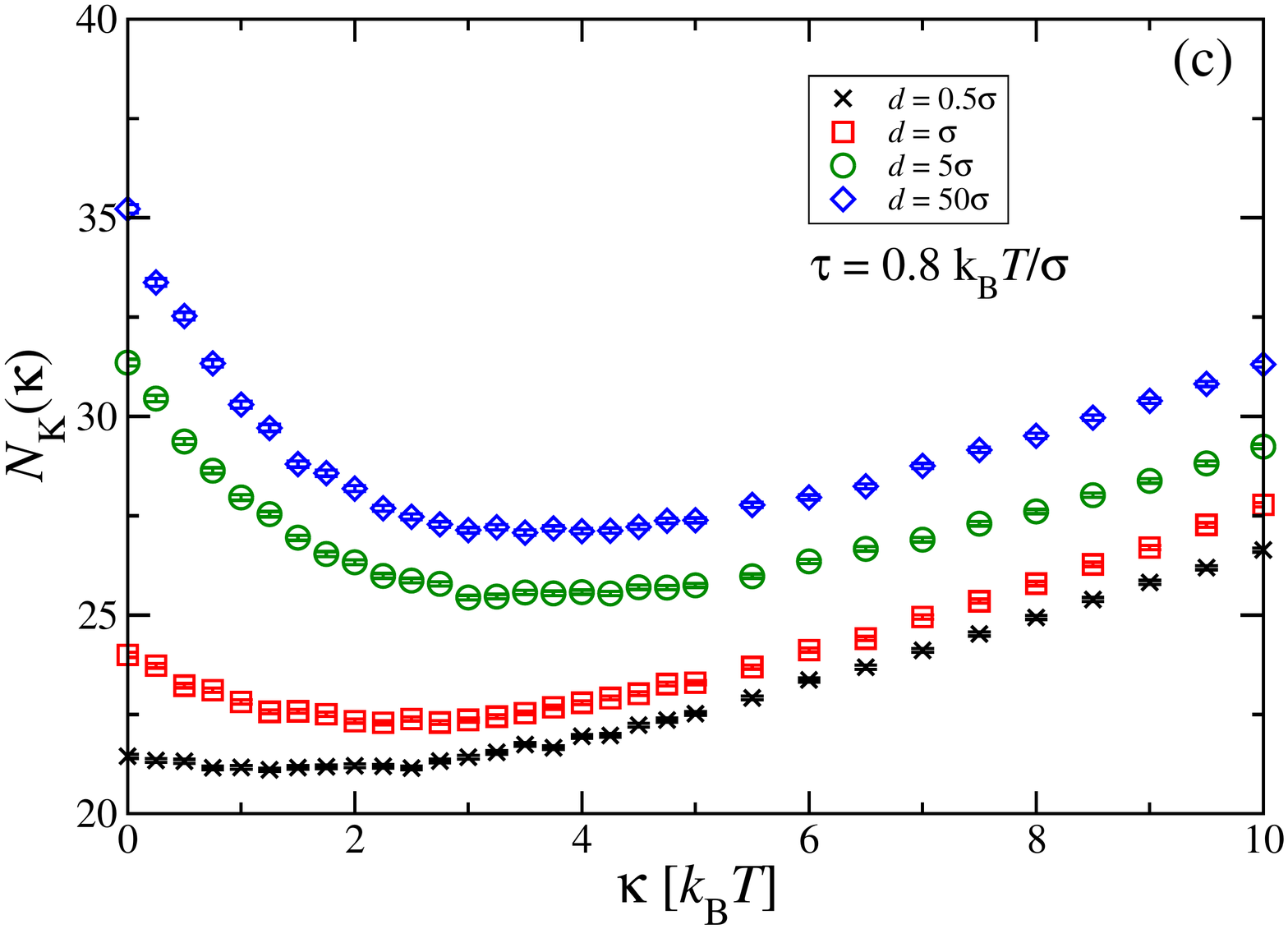} 
}
{
\includegraphics[width=7.5cm]{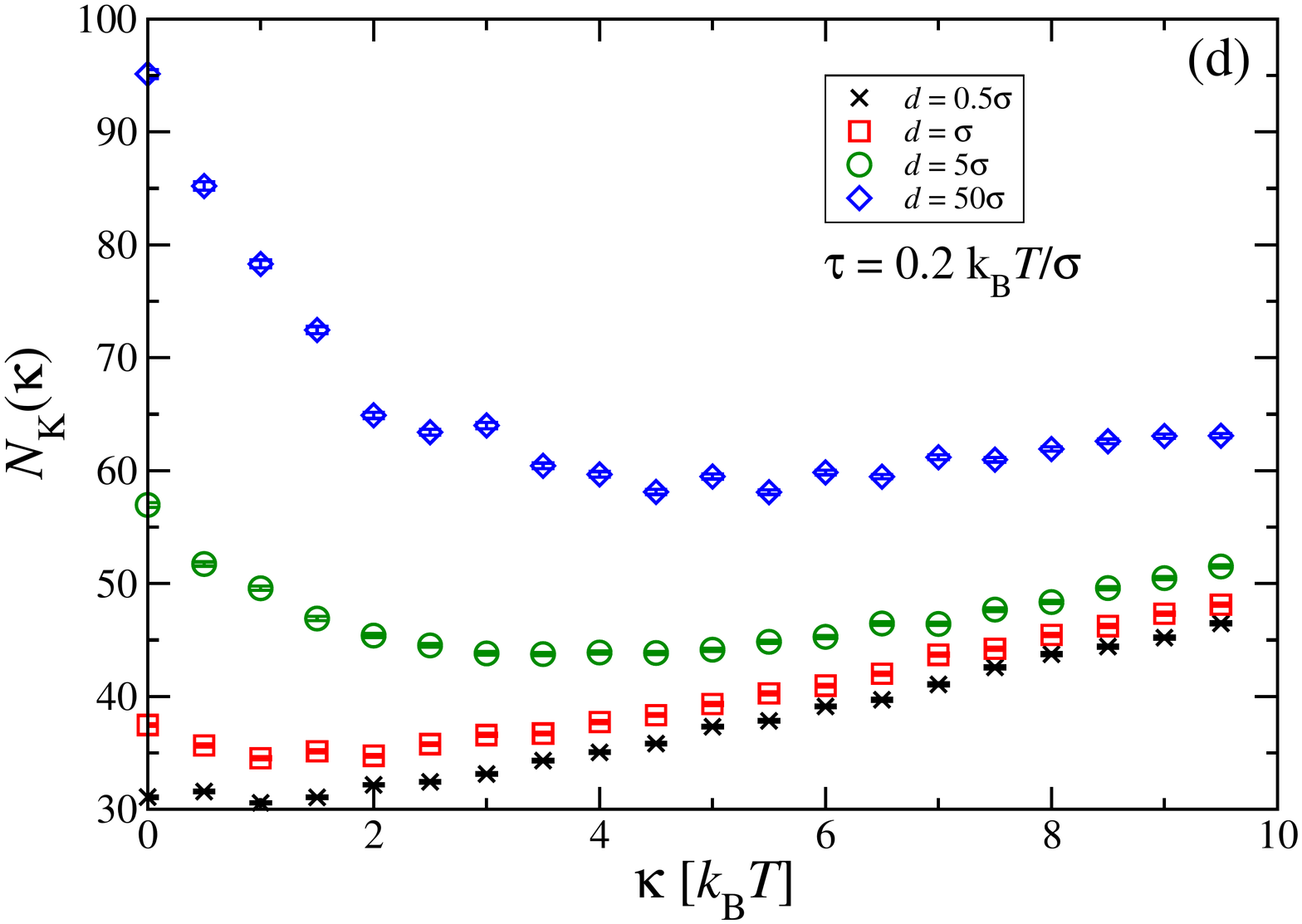} 
}
}
\caption{(a), (b): The quantity $\Psi(\kappa)$ for different confinements, 
as indicated in the legend. The inset in (a) shows a zoom of the main panel in the
region of the minimum of $\Psi(\kappa)$. (c), (d) The dependence of the number of monomers in the knot, $N_{\rm K}(\kappa)$, on chain rigidity. 
In panels (a) and (c) results for the tension $\tau = 0.8\,k_{\rm B}T/\sigma$ are shown, 
whereas in panels (b) and (d) the tension is $\tau = 0.2\,k_{\rm B}T/\sigma$.}
\label{FKappaPlot}
\end{figure}

We now turn our attention to the confined case. It turns out that
the effects of confinement are most transparent at a small, but non-zero value of the bending rigidity, thus we show in
Figs.~\ref{AddBendingPerDeltaPlots}(c) and \ref{AddBendingPerDeltaPlots}(d)
results for $\Delta b_{\alpha}(\kappa)$ for $\beta\kappa = 0.5$, which is characteristic for all values
of $\kappa \leq \kappa_{\rm min}$, the latter being the value of the rigidity for which $\Psi(\kappa)$ attains its minimum value $\Psi_{\rm min}$. 
Here, striking differences between the tensions $\tau = 0.8\,k_{\rm B}T/\sigma$ and $\tau = 0.1\,k_{\rm B}T/\sigma$ show up. 
While for $\tau = 0.8\,k_{\rm B}T/\sigma$ there is hardly any difference between the confined and
unconfined polymers for slit widths as small as $d = 5\sigma$, for $\tau = 0.1\,k_{\rm B}T/\sigma$, confinement enhances $\Delta b_{\alpha}$ at the beginning 
of the knot by almost a factor two. For both tensions, the suppression of
bending through knotting becomes {\it even stronger} as a result of the geometric constraints and thus $\Delta B(\kappa)$ 
is more negative for the polymer in the slit than it is for the free polymer. This is consistent with the interpretation given above
for the case of the unconfined polymer, since the slit confinement forces the strands at the braid region to come closer which further reduces the random bending in the braid region. This effect, however, is more pronounced for smaller tensions, 
where the braiding region without confinement is looser than for polymer chains under higher tensions.

There exists a correlation
between the slopes of $\Psi(\kappa)$ and $N_{\rm K}(\kappa)$
shown in Fig.\ \ref{FKappaPlot}. 
In the high-$\kappa$ domain, $\Psi(\kappa) \sim N_{\rm K}(\kappa) \sim \sqrt{\kappa}$,
as is evident from the discussion following Eq.\ (\ref{integrate:eq}).
For $\kappa = 0$,  the knot is swollen due to the presence of steric interactions, maximizing in this way
its entropy. However, for non-zero $\kappa \sim k_{\rm B}T$, the fluctuations of the monomers are restricted in the first place, 
enabling thus a tighter braided region and a concomitant reduction of knot size. Thus also for small $\kappa$ 
the slopes of $\Psi(\kappa)$ and $N_{\rm K}(\kappa)$ are expected to have the same sign.
Note, however, that the value $\tilde\kappa$ that minimizes $N_{\rm K}(\kappa)$ does not coincide with $\kappa_{\rm min}$, e.g., for very strong confinements
$\kappa_{\rm min} \ne 0$, whereas $\tilde\kappa$ is, within simulation resolution, vanishingly small.

\begin{figure}[htp]
\subfigure{
{
\includegraphics[width=7.5cm]{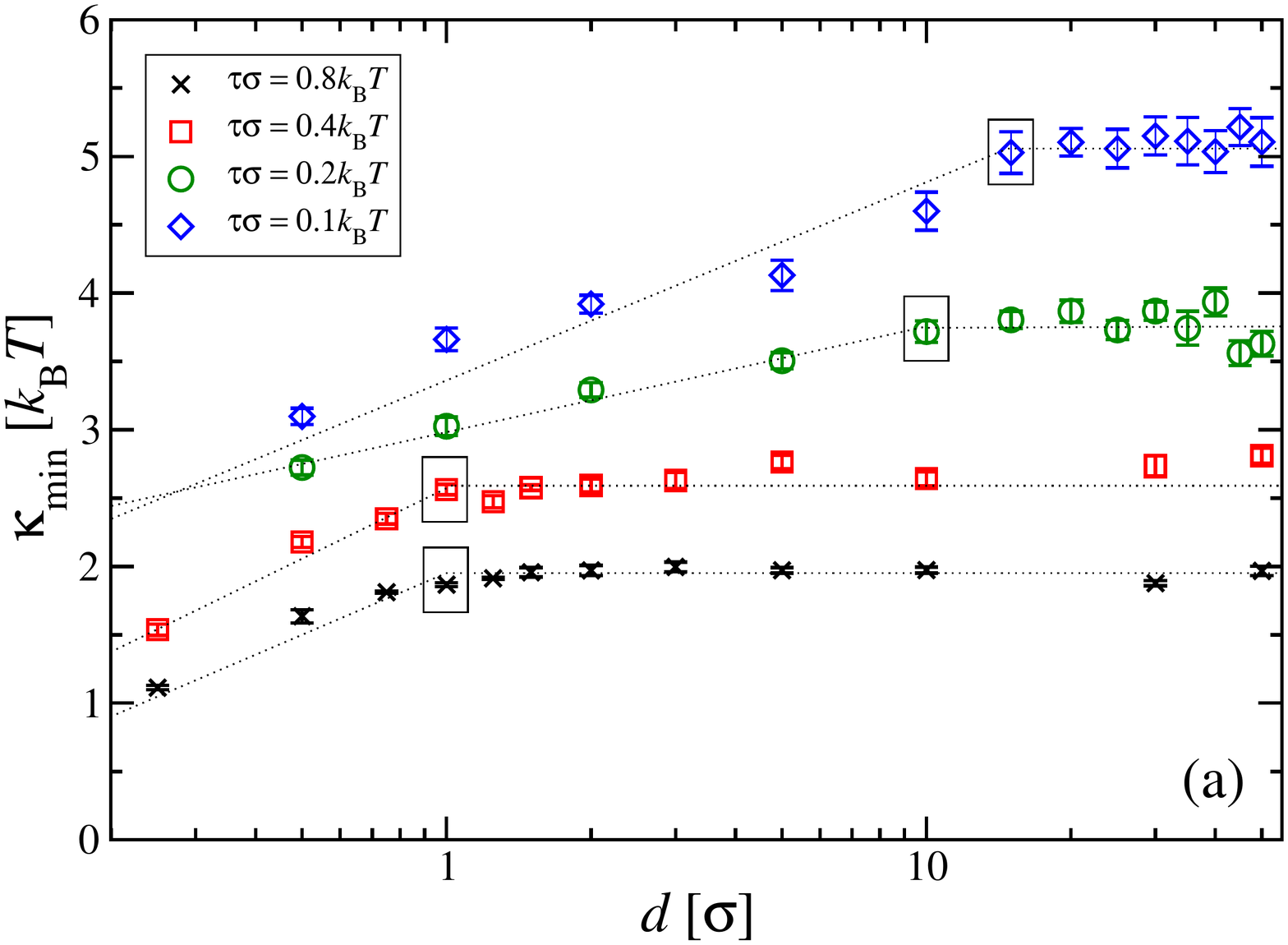} 
}
\vspace{0.3cm}
}
\subfigure{
{
\includegraphics[width=7.5cm]{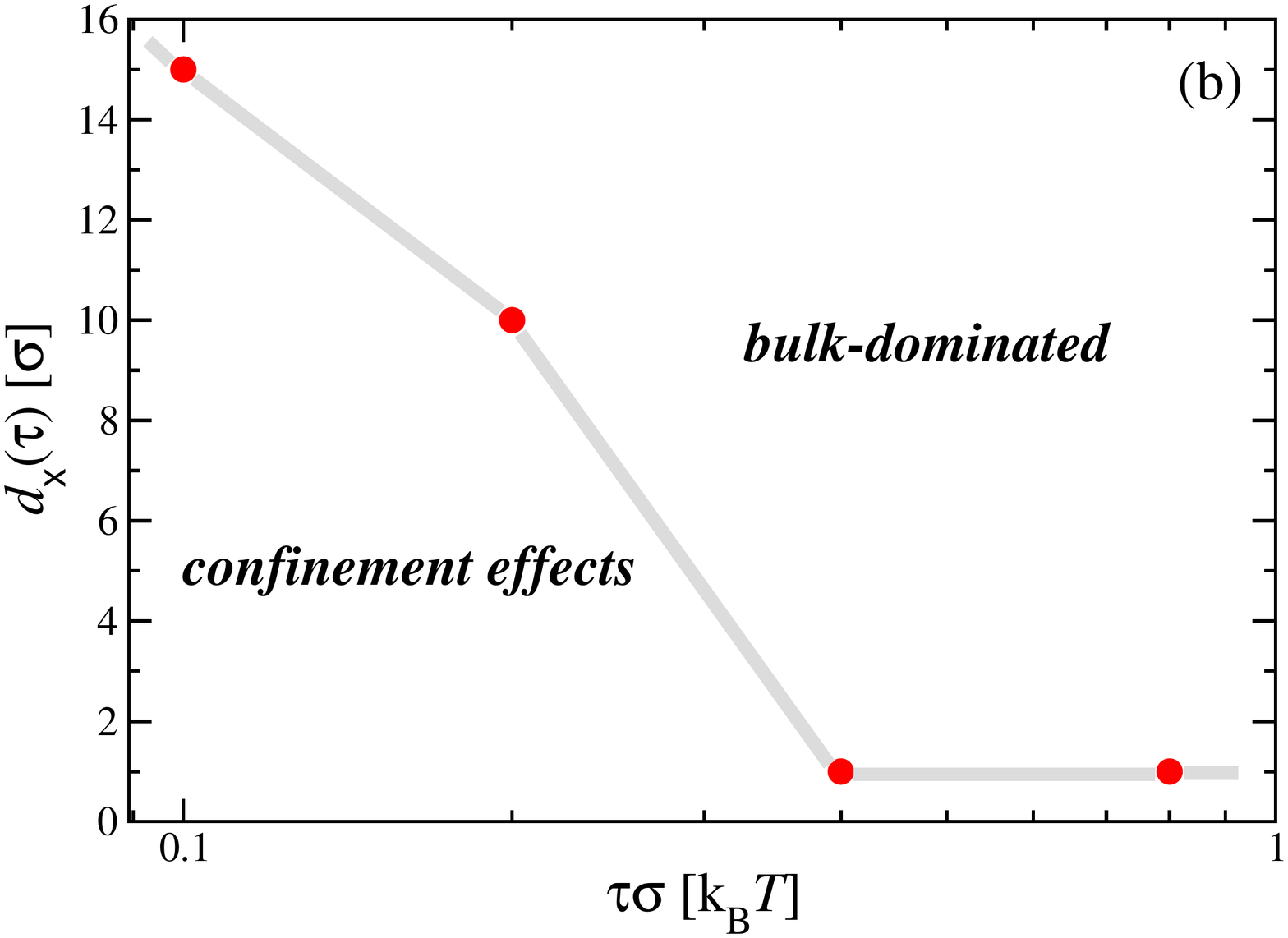} 
}
}
\caption{(a): The dependence of the optimal value for the rigidity, $\kappa_{\rm min}$, on confinement for several applied tensions. 
The dotted lines are guides to the eye, delineating the two regimes where $\kappa_{\rm min}$ is influenced by confinement, where 
$\kappa_{\rm min}$ is $d$-dependent, and the 
and the bulk-dominated regime, where it is not. The crossover data points $d_{\times}(\tau)$ 
between the two regimes are marked with boxes. 
(b) The dependence of $d_{\times}(\tau)$ on the applied tension $\tau$. The data points (red circles) are connected
with thick gray segments, separating the bulk-dominated regime above the line by the confinement-affected regime below it.}
\label{TauKOptPlots}
\end{figure}

\begin{figure}[htp]
\subfigure{
{
\includegraphics[width=7.5cm]{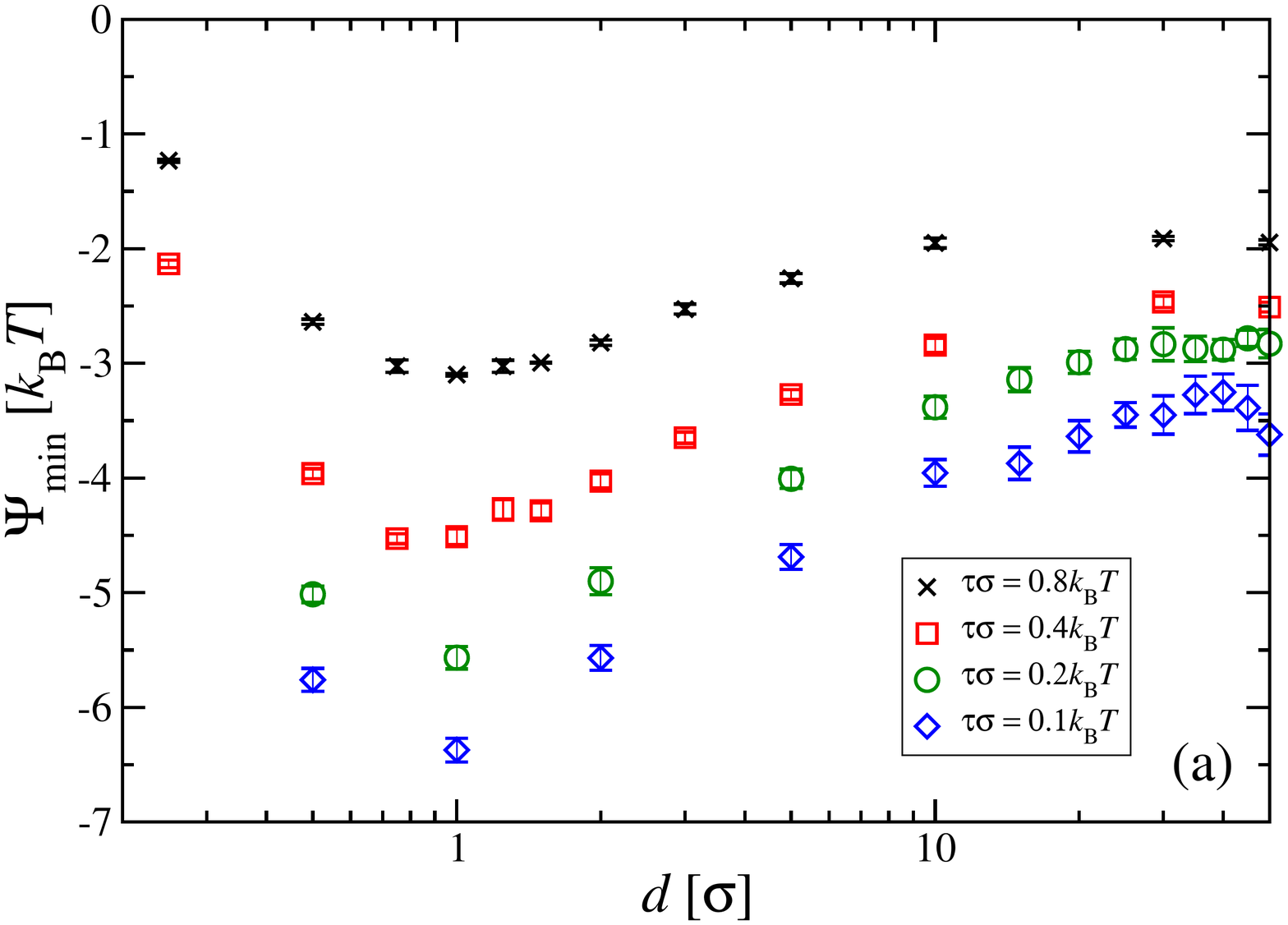} 
}
\vspace{0.3cm}
}
\subfigure{
{
\includegraphics[width=7.5cm]{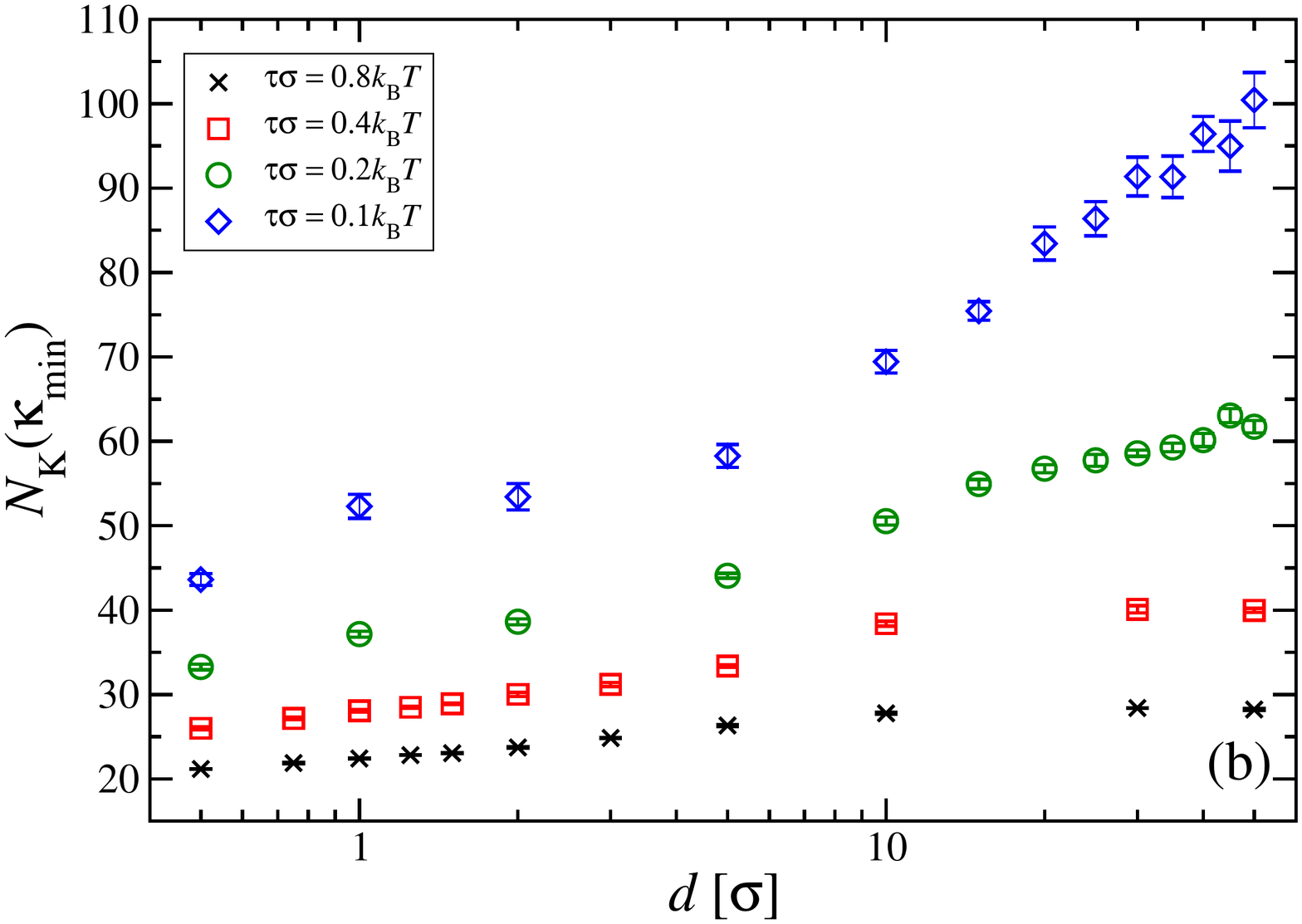} 
}
} 
\caption{(a): The dependence of $\Psi_{\min}$, the value of $\Psi(\kappa)$ at the optimal rigidity $\kappa_{\rm min}$, on $d$, the slit-with of the confinement. 
(b): The knot size at the optimal rigidity as a function of $d$. Applied tensions as indicated in the legends.}
\label{TauFNKPlots}
\end{figure}

The dependence of the rigidity $\kappa_{\rm min}$ for which  $\Psi(\kappa)$ has its minimum
on the degree of confinement for different applied tensions is summarized in 
Fig.~\ref{TauKOptPlots}(a).
We find that
for a tension of $\tau \sigma= 0.8 k_{\rm B}T$, $\kappa_{\rm min}$ is only affected by
confinement for slit-widths lying at the monomer scale. In this 
regime of ultra-strong confinement, the energy cost that chain segments would have to pay to go one above the other in a gradual fashion at the 
braiding regions are too high. This is caused by the external potential, which assigns an increasingly high energetic cost 
for every monomer that deviates strongly from the $y = 0$-plane.
Accordingly, it is preferable for the system to form localized `kinks' of one or two
monomers in the braiding region, which expose a minimal number of monomers to the regions of high external potential, while
at the same time creating strong bending there. 

The situation is quite different, however, for lower tensions.
In this case,
also a moderate confinement of the order of 10 bond lengths, can significantly affect the value of $\kappa_{\rm min}$. 
To better quantify the effects of confinement, we employ a simple, rough-and-ready separation of the data points shown
in Fig.~\ref{TauKOptPlots}(a) into two groups: for high values of $d$, the points form plateaus at the bulk values of $\kappa_{\rm min}$,
which we connect by horizontal lines. Through the other groups of points straight lines are drawn by hand, which intersect the horizontal ones at 
tension-dependent crossover confinement widths $d_{\times}(\tau)$. These values denote, by construction, the crossover of the
behaviour of $\kappa_{\rm min}$ from bulk-dominated, for $d > d_{\times}(\tau)$, to confinement-affected, for $d < d_{\times}(\tau)$.
The results are summarized in 
Fig.~\ref{TauKOptPlots}(b), 
where it can be seen that
$d_{\times}(\tau)$ is significantly increased for lower tensions. 
As we discuss below, the reason the situation is strikingly different for lower tensions seems to be related to the fact that the knot size is then significantly increased with respect to higher tensions.

The increased effects of confinement as $\tau$ decreases are also manifested on the value of $\Psi_{\rm min}$ as well as on the knot size
$N_K(\kappa_{\rm min})$. The former quantity is shown in Fig.~\ref{TauFNKPlots}(a) and the latter in Fig.~\ref{TauFNKPlots}(b).
As can be seen in Fig.~\ref{TauFNKPlots}(a), and in contrast to the values $\kappa_{\rm min}$ itself, even in the case of higher 
tension the corresponding depth of the minimum $\Psi_{\rm min}$ is influenced by a confining slit width of the order of 10 bond lengths. 
However, for lower tensions, the effect on $\Psi_{\rm min}$ is felt at
even larger slit widths. Furthermore, the difference between the value $\Psi_{\rm min}$ in the bulk ($d/\sigma \gg 1$) 
and the one for the optimal slit
width is enhanced. The fact that the effect of the confinement on the knot size is more pronounced for smaller tensions, as shown in Fig.~\ref{TauFNKPlots}(b), correlates well with the finding that confinement shifts $\kappa_{\rm min}$ to lower values for sufficiently small tensions.
As we have seen in Figs.~\ref{AddBendingPerDeltaPlots}(a) and \ref{AddBendingPerDeltaPlots}(b) 
one of the contributions that eventually render $\partial\Psi(\kappa)/\partial\kappa$ positive, 
is the bending in the interior of the knot, 
which arises as the knot enforces the polymer to form a loop. This contribution is larger for smaller knot sizes, and it is therefore consistent
that $\kappa_{\rm min}$ will be shifted by confinement if the latter is able to significantly reduce the knot size. 

\begin{figure}[htp]
\subfigure{
\includegraphics[width=7.5cm]{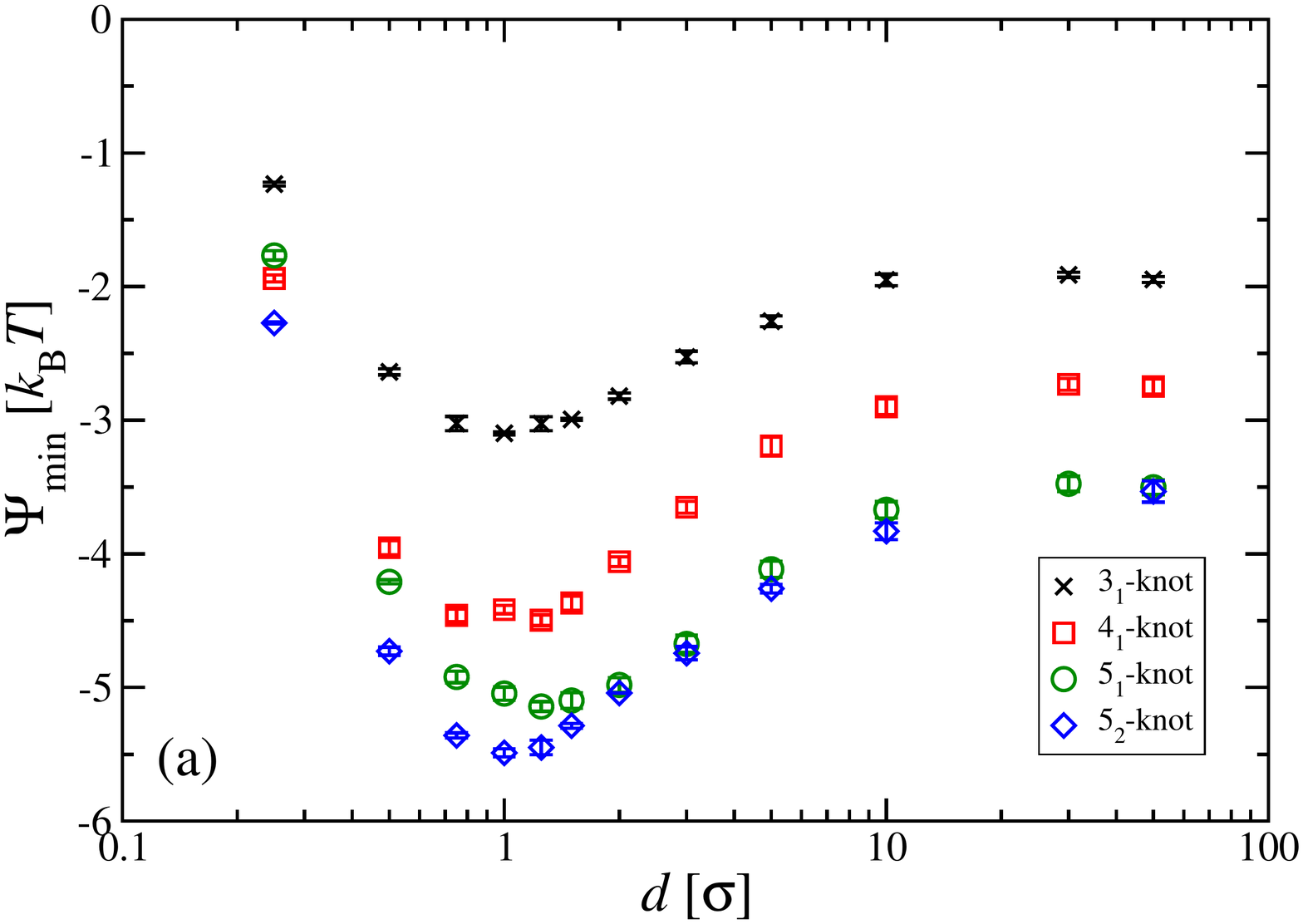} 
}
\subfigure{
\includegraphics[width=7.5cm]{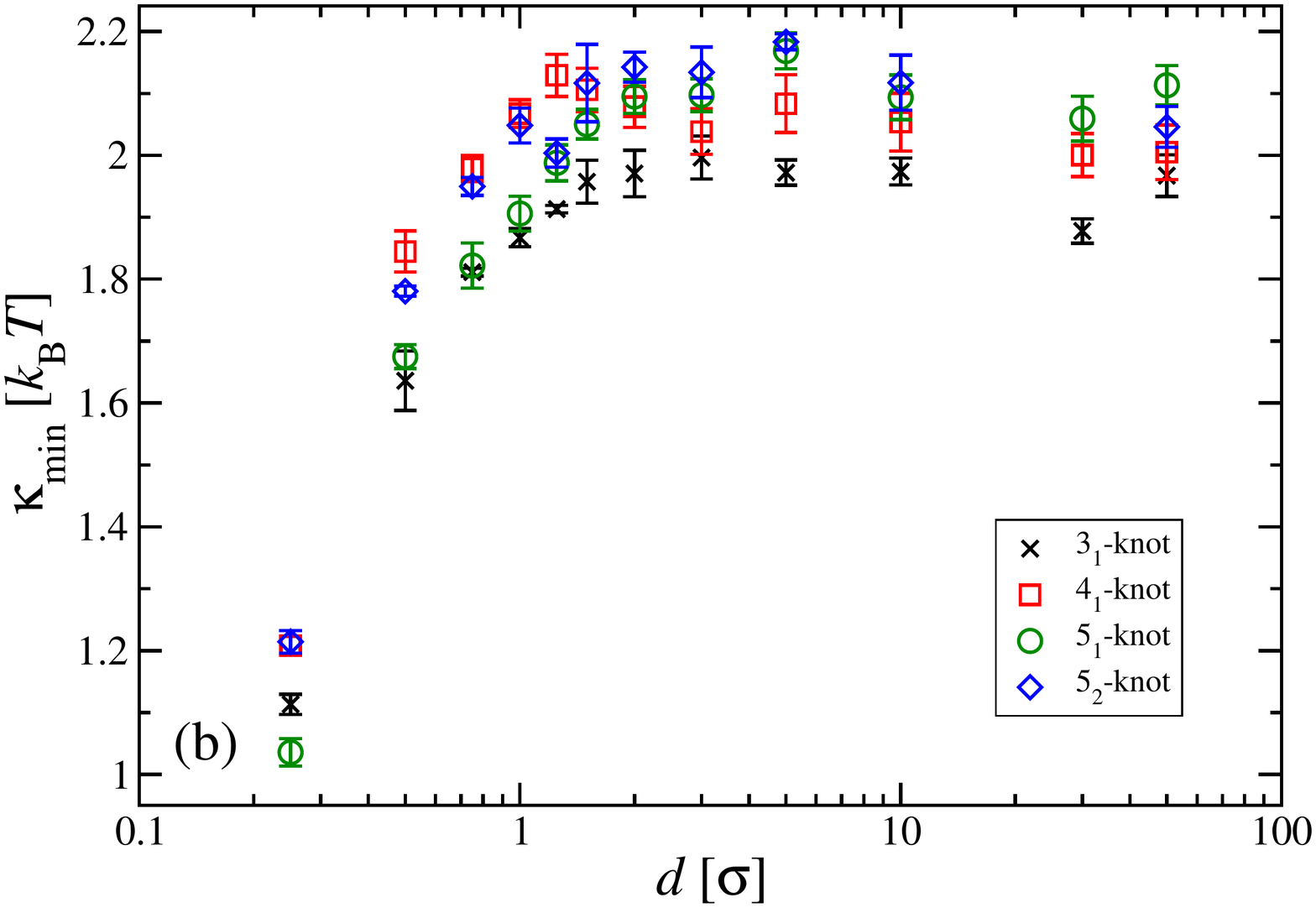} 
}
\caption{(a) The dependence of $\Psi_{\rm min}$ on confinement for tension $\tau=0.8\,k_{\rm B}T/\sigma$ and different knot types,
as indicated in the legend. (b) The corresponding optimal value of the rigidity $\kappa_{\rm min}$ for different slit widths and knot types.}
\label{KTOptPlot}
\end{figure}

Up to now, all results have been derived for the simplest, trefoil knot; real polymers can, however, display a large variety of increasingly 
complex knots \cite{koniaris:prl:1991,rawdon:mm:2008}. Considering other knots allows us on the one hand to put the general character of our results to the test, and also to corroborate our assertion that the crossings at the braiding region are responsible for the reduction of $\Psi(\kappa)$ at
finite $\kappa$-values. Indeed, more complex knots have more crossing points, where the strands of the polymer chain interact with each other. 
Thus, according to our analysis above, one should expect that a more complex knot will lead to lower values of $\Delta B(\kappa)$ 
and $\Psi_{\rm min}$ for small but finite $\kappa$. Our findings for different knot topologies 
(denoted in the Alexander-Briggs notation \cite{livingston1993knot}) are summarized in Figs.\ \ref{KTOptPlot}(a)
and \ref{KTOptPlot}(b).
The data in Fig.\ \ref{KTOptPlot}(a) confirm that the main effect of the increased knot complexity 
is the addition of crossing points which all result in a similar bending suppression as the crossing points of the trefoil knot. Accordingly, for the knot topologies investigated,
$\Psi_{\rm min}$ is approximately proportional to the number of minimal crossings of the respective knot diagram, 
at least for confinements with $d \geq \sigma$. It is also striking that for the unconfined case, $\Psi_{\rm min}$ 
is, within error bars, identical for the $5_1$ and $5_2$ topologies. A test of whether $\Psi_{\rm min}$ 
is in good approximation proportional to the minimal number of crossings of arbitrarily complex knots is beyond the scope of this work. 
However, the number of strand-crossings will increase with knot complexity. Accordingly, the free energy penalty for putting a knot on a stiff polymer
($\kappa \ne 0$) can be much lower than the one for putting it on a fully flexible polymer ($\kappa = 0$), by amounts that grow
with the knot complexity. Whereas $\Psi_{\rm min}$ is sensitive to the knot type, $\kappa_{\rm min}$ is not, as can be ascertained from the results shown
in Fig.\ \ref{KTOptPlot}(b). All data fall within a narrow band of width $\Delta\kappa_{\rm min} \cong 0.2 k_{\rm B}T$ irrespective of the knot topology.

The tensions considered in our work are of the order of $k_B T/\sigma$. 
At room temperature and a monomer length scale of $1\,{\rm nm}$ this corresponds to the ${\rm pN}$-scale. 
These tensions result in values of $\kappa_{\rm min} \approx 5\,k_{\rm B}T$. As it was found in previous work \cite{Matthews2012}, 
without confinement $\kappa_{\rm min}$ scales approximately as $\sim {\tau}^{-1/2}$. A reduction of the
tension down to the ${\rm fN}$-scale, which is typical of double-stranded
DNA molecules \cite{lipowsky:epl:2006}, will bring $\kappa_{\rm min}$ at the order of $100\,k_{\rm B}T$. Our results imply that at these 
lower tensions the effect of confinement on $\kappa_{\rm min}$ can be expected to be even more pronounced.

\section{Conclusions}
\label{sec:conclusions}
In summary, we have demonstrated that the local stretching at the braiding region and close to the crossing points is the physical mechanism
responsible for the minimization of the free energy penalty of knotting of a linear polymer for non-vanishing values of the bending rigidity.
Confinement can affect the location of the optimal rigidity for sufficiently low tensions, when it is at the same time significantly affecting the knot size. 
We therefore expect the geometrical reduction of dimensionality to become
relevant for the location of the knots of chains with variable rigidity if the latter are under sufficiently small tensions. 
For tensions at the fN-scale, which are typical of double-stranded DNA molecules\cite{lipowsky:epl:2006}, 
we therefore expect that confinement to strongly influence the value of the optimal rigidity.
On the other hand, the amount of reduction of the knotting of free energy by rigidity strongly depends on the topology of the knot and it
increases with the knot complexity, scaling roughly with the number of minimal crossings of the knot. Accordingly, we anticipate that more
complex knots will localize more strongly in the optimal regions of a chain than simpler ones. 
Recent advances in tying knots on polymers by optical tweezers \cite{arai:nature:1999,bao:prl:2003} 
and adsorbing them on mica surfaces \cite{ercolini:prl:2007} should
allow for experimental testing of our predictions.

\begin{acknowledgement}
This work has been supported by the Austrian Science Fund (FWF), Grant 23400-N16.
\end{acknowledgement}

\bibliography{bib_knots.bib}

\providecommand*\mcitethebibliography{\thebibliography}
\csname @ifundefined\endcsname{endmcitethebibliography}
  {\let\endmcitethebibliography\endthebibliography}{}
\begin{mcitethebibliography}{31}
\providecommand*\natexlab[1]{#1}
\providecommand*\mciteSetBstSublistMode[1]{}
\providecommand*\mciteSetBstMaxWidthForm[2]{}
\providecommand*\mciteBstWouldAddEndPuncttrue
  {\def\EndOfBibitem{\unskip.}}
\providecommand*\mciteBstWouldAddEndPunctfalse
  {\let\EndOfBibitem\relax}
\providecommand*\mciteSetBstMidEndSepPunct[3]{}
\providecommand*\mciteSetBstSublistLabelBeginEnd[3]{}
\providecommand*\EndOfBibitem{}
\mciteSetBstSublistMode{f}
\mciteSetBstMaxWidthForm{subitem}{(\alph{mcitesubitemcount})}
\mciteSetBstSublistLabelBeginEnd
  {\mcitemaxwidthsubitemform\space}
  {\relax}
  {\relax}

\bibitem[Matthews et~al.(2012)Matthews, Louis, and Likos]{Matthews2012}
Matthews,~R.; Louis,~A.~A.; Likos,~C.~N. \emph{ACS Macro Letters}
  \textbf{2012}, \emph{1}, 1352\relax
\mciteBstWouldAddEndPuncttrue
\mciteSetBstMidEndSepPunct{\mcitedefaultmidpunct}
{\mcitedefaultendpunct}{\mcitedefaultseppunct}\relax
\EndOfBibitem
\bibitem[Sogo et~al.(1999)Sogo, Stasiak, Mart{\'{\i}}nobles, Krimer,
  Hern{\'a}ndez, and Schvartzman]{Sogo1999}
Sogo,~J.~M.; Stasiak,~A.; Mart{\'{\i}}nobles,~M.~L.; Krimer,~D.~B.;
  Hern{\'a}ndez,~P.; Schvartzman,~J.~B. \emph{J. Mol. Biol.} \textbf{1999},
  \emph{286}, 637\relax
\mciteBstWouldAddEndPuncttrue
\mciteSetBstMidEndSepPunct{\mcitedefaultmidpunct}
{\mcitedefaultendpunct}{\mcitedefaultseppunct}\relax
\EndOfBibitem
\bibitem[Arsuaga et~al.(2002)Arsuaga, V{\'a}zquez, Trigueros, Sumners, and
  Roca]{Arsuaga2002}
Arsuaga,~J.; V{\'a}zquez,~M.; Trigueros,~S.; Sumners,~D.~W.; Roca,~J.
  \emph{Proc. Nat. Acad. Sci. U.S.A.} \textbf{2002}, \emph{99}, 5373\relax
\mciteBstWouldAddEndPuncttrue
\mciteSetBstMidEndSepPunct{\mcitedefaultmidpunct}
{\mcitedefaultendpunct}{\mcitedefaultseppunct}\relax
\EndOfBibitem
\bibitem[Portugal and Rodr{\'\i}guez-Campos(1996)Portugal, and
  Rodr{\'\i}guez-Campos]{Portugal1996}
Portugal,~J.; Rodr{\'\i}guez-Campos,~A. \emph{Nucleic Acids Res.}
  \textbf{1996}, \emph{24}, 4890\relax
\mciteBstWouldAddEndPuncttrue
\mciteSetBstMidEndSepPunct{\mcitedefaultmidpunct}
{\mcitedefaultendpunct}{\mcitedefaultseppunct}\relax
\EndOfBibitem
\bibitem[Deibler et~al.(2007)Deibler, Mann, {De Witt L. Sumners}, and
  Zechiedrich]{Deibler2007}
Deibler,~R.~W.; Mann,~J.~K.; {De Witt L. Sumners},; Zechiedrich,~L. \emph{BMC
  Mol. Biol.} \textbf{2007}, \emph{8}, 44\relax
\mciteBstWouldAddEndPuncttrue
\mciteSetBstMidEndSepPunct{\mcitedefaultmidpunct}
{\mcitedefaultendpunct}{\mcitedefaultseppunct}\relax
\EndOfBibitem
\bibitem[Liu et~al.(2009)Liu, Deibler, Chan, and Zechiedrich]{Liu2009}
Liu,~Z.; Deibler,~R.~W.; Chan,~H.~S.; Zechiedrich,~L. \emph{Nucleic Acids Res.}
  \textbf{2009}, \emph{37}, 661\relax
\mciteBstWouldAddEndPuncttrue
\mciteSetBstMidEndSepPunct{\mcitedefaultmidpunct}
{\mcitedefaultendpunct}{\mcitedefaultseppunct}\relax
\EndOfBibitem
\bibitem[Hogan et~al.(1983)Hogan, LeGrange, and Austin]{Hogan1983}
Hogan,~M.; LeGrange,~J.; Austin,~B. \emph{Nature} \textbf{1983}, \emph{304},
  752\relax
\mciteBstWouldAddEndPuncttrue
\mciteSetBstMidEndSepPunct{\mcitedefaultmidpunct}
{\mcitedefaultendpunct}{\mcitedefaultseppunct}\relax
\EndOfBibitem
\bibitem[Geggier and Vologodskii(2010)Geggier, and Vologodskii]{Geggier2010}
Geggier,~S.; Vologodskii,~A. \emph{Proc. Nat. Acad. Sci. U.S.A.} \textbf{2010},
  \emph{107}, 15421\relax
\mciteBstWouldAddEndPuncttrue
\mciteSetBstMidEndSepPunct{\mcitedefaultmidpunct}
{\mcitedefaultendpunct}{\mcitedefaultseppunct}\relax
\EndOfBibitem
\bibitem[Johnson et~al.(2013)Johnson, Chen, and Phillips]{Johnson2013}
Johnson,~S.; Chen,~Y.-J.; Phillips,~R. \emph{PLoS ONE} \textbf{2013}, \emph{8},
  e75799\relax
\mciteBstWouldAddEndPuncttrue
\mciteSetBstMidEndSepPunct{\mcitedefaultmidpunct}
{\mcitedefaultendpunct}{\mcitedefaultseppunct}\relax
\EndOfBibitem
\bibitem[Rybenkov(1997)]{Rybenkov1997}
Rybenkov,~V.~V. \emph{Science} \textbf{1997}, \emph{277}, 690\relax
\mciteBstWouldAddEndPuncttrue
\mciteSetBstMidEndSepPunct{\mcitedefaultmidpunct}
{\mcitedefaultendpunct}{\mcitedefaultseppunct}\relax
\EndOfBibitem
\bibitem[Emanuel et~al.(2009)Emanuel, {Hamedani Radja}, Henriksson, and
  Schiessel]{Emanuel2009}
Emanuel,~M.; {Hamedani Radja},~N.; Henriksson,~A.; Schiessel,~H. \emph{Phys.
  Biol.} \textbf{2009}, \emph{6}, 025008\relax
\mciteBstWouldAddEndPuncttrue
\mciteSetBstMidEndSepPunct{\mcitedefaultmidpunct}
{\mcitedefaultendpunct}{\mcitedefaultseppunct}\relax
\EndOfBibitem
\bibitem[Jun and Mulder(2006)Jun, and Mulder]{Jun2006}
Jun,~S.; Mulder,~B. \emph{Proc. Nat. Acad. Sci. U.S.A.} \textbf{2006},
  \emph{103}, 12388\relax
\mciteBstWouldAddEndPuncttrue
\mciteSetBstMidEndSepPunct{\mcitedefaultmidpunct}
{\mcitedefaultendpunct}{\mcitedefaultseppunct}\relax
\EndOfBibitem
\bibitem[Purohit et~al.(2005)Purohit, Inamdar, Grayson, Squires, Kondev, and
  Phillips]{Purohit2005}
Purohit,~P.~K.; Inamdar,~M.~M.; Grayson,~P.~D.; Squires,~T.~M.; Kondev,~J.;
  Phillips,~R. \emph{Biophys. J.} \textbf{2005}, \emph{88}, 851\relax
\mciteBstWouldAddEndPuncttrue
\mciteSetBstMidEndSepPunct{\mcitedefaultmidpunct}
{\mcitedefaultendpunct}{\mcitedefaultseppunct}\relax
\EndOfBibitem
\bibitem[Orlandini et~al.(2009)Orlandini, Stella, and
  Vanderzande]{orlandini2009}
Orlandini,~E.; Stella,~A.~L.; Vanderzande,~C. \emph{Phys. Biol.} \textbf{2009},
  \emph{6}, 025012\relax
\mciteBstWouldAddEndPuncttrue
\mciteSetBstMidEndSepPunct{\mcitedefaultmidpunct}
{\mcitedefaultendpunct}{\mcitedefaultseppunct}\relax
\EndOfBibitem
\bibitem[Ercolini et~al.(2007)Ercolini, Valle, Adamcik, Witz, Metzler, {de los
  Rios}, Roca, and Dietler]{ercolini:prl:2007}
Ercolini,~E.; Valle,~F.; Adamcik,~J.; Witz,~G.; Metzler,~R.; {de los Rios},~P.;
  Roca,~J.; Dietler,~G. \emph{Phys. Rev. Lett.} \textbf{2007}, \emph{98},
  058102\relax
\mciteBstWouldAddEndPuncttrue
\mciteSetBstMidEndSepPunct{\mcitedefaultmidpunct}
{\mcitedefaultendpunct}{\mcitedefaultseppunct}\relax
\EndOfBibitem
\bibitem[Micheletti and Orlandini(2012)Micheletti, and
  Orlandini]{Micheletti2012}
Micheletti,~C.; Orlandini,~E. \emph{Macromolecules} \textbf{2012}, \emph{45},
  2113\relax
\mciteBstWouldAddEndPuncttrue
\mciteSetBstMidEndSepPunct{\mcitedefaultmidpunct}
{\mcitedefaultendpunct}{\mcitedefaultseppunct}\relax
\EndOfBibitem
\bibitem[Matthews et~al.(2011)Matthews, Louis, and Yeomans]{Matthews2011}
Matthews,~R.; Louis,~A.~A.; Yeomans,~J.~M. \emph{Mol. Phys.} \textbf{2011},
  \emph{109}, 1289\relax
\mciteBstWouldAddEndPuncttrue
\mciteSetBstMidEndSepPunct{\mcitedefaultmidpunct}
{\mcitedefaultendpunct}{\mcitedefaultseppunct}\relax
\EndOfBibitem
\bibitem[Plimpton(1995)]{Plimpton19951}
Plimpton,~S. \emph{J. Comput. Phys.} \textbf{1995}, \emph{117}, 1\relax
\mciteBstWouldAddEndPuncttrue
\mciteSetBstMidEndSepPunct{\mcitedefaultmidpunct}
{\mcitedefaultendpunct}{\mcitedefaultseppunct}\relax
\EndOfBibitem
\bibitem[Tuckerman et~al.(2001)Tuckerman, Liu, Ciccotti, and
  Martyna]{Tuckerman2001}
Tuckerman,~M.~E.; Liu,~Y.; Ciccotti,~G.; Martyna,~G.~J. \emph{J. Chem. Phys.}
  \textbf{2001}, \emph{115}, 1678\relax
\mciteBstWouldAddEndPuncttrue
\mciteSetBstMidEndSepPunct{\mcitedefaultmidpunct}
{\mcitedefaultendpunct}{\mcitedefaultseppunct}\relax
\EndOfBibitem
\bibitem[Rubinstein and Colby(2003)Rubinstein, and
  Colby]{rubinstein2003polymer}
Rubinstein,~M.; Colby,~R. \emph{Polymer Physics}; OUP Oxford, 2003\relax
\mciteBstWouldAddEndPuncttrue
\mciteSetBstMidEndSepPunct{\mcitedefaultmidpunct}
{\mcitedefaultendpunct}{\mcitedefaultseppunct}\relax
\EndOfBibitem
\bibitem[Gallotti and Pierre-Louis(2007)Gallotti, and
  Pierre-Louis]{Gallotti2007}
Gallotti,~R.; Pierre-Louis,~O. \emph{Phys. Rev. E} \textbf{2007}, \emph{75},
  031801\relax
\mciteBstWouldAddEndPuncttrue
\mciteSetBstMidEndSepPunct{\mcitedefaultmidpunct}
{\mcitedefaultendpunct}{\mcitedefaultseppunct}\relax
\EndOfBibitem
\bibitem[Tubiana et~al.(2011)Tubiana, Orlandini, and Micheletti]{Tubiana2011}
Tubiana,~L.; Orlandini,~E.; Micheletti,~C. \emph{Progr. Theor. Phys. Supp.}
  \textbf{2011}, \emph{191}, 192\relax
\mciteBstWouldAddEndPuncttrue
\mciteSetBstMidEndSepPunct{\mcitedefaultmidpunct}
{\mcitedefaultendpunct}{\mcitedefaultseppunct}\relax
\EndOfBibitem
\bibitem[Marcone et~al.(2007)Marcone, Orlandini, Stella, and
  Zonta]{Marcone2007}
Marcone,~B.; Orlandini,~E.; Stella,~A.~L.; Zonta,~F. \emph{Phys. Rev. E}
  \textbf{2007}, \emph{75}, 041105\relax
\mciteBstWouldAddEndPuncttrue
\mciteSetBstMidEndSepPunct{\mcitedefaultmidpunct}
{\mcitedefaultendpunct}{\mcitedefaultseppunct}\relax
\EndOfBibitem
\bibitem[Marcone et~al.(2005)Marcone, Orlandini, Stella, and
  Zonta]{Marcone2005}
Marcone,~B.; Orlandini,~E.; Stella,~A.~L.; Zonta,~F. \emph{J. Phys. A: Math.
  Gen.} \textbf{2005}, \emph{38}, L15\relax
\mciteBstWouldAddEndPuncttrue
\mciteSetBstMidEndSepPunct{\mcitedefaultmidpunct}
{\mcitedefaultendpunct}{\mcitedefaultseppunct}\relax
\EndOfBibitem
\bibitem[Koniaris and Muthukumar(1991)Koniaris, and
  Muthukumar]{koniaris:prl:1991}
Koniaris,~K.; Muthukumar,~M. \emph{Phys. Rev. Lett.} \textbf{1991}, \emph{66},
  2211\relax
\mciteBstWouldAddEndPuncttrue
\mciteSetBstMidEndSepPunct{\mcitedefaultmidpunct}
{\mcitedefaultendpunct}{\mcitedefaultseppunct}\relax
\EndOfBibitem
\bibitem[Rawdon et~al.(2008)Rawdon, Dobay, Kern, Millett, Piatek, Plunkett, and
  Stasiak]{rawdon:mm:2008}
Rawdon,~E.; Dobay,~A.; Kern,~J.~C.; Millett,~K.~C.; Piatek,~M.; Plunkett,~P.;
  Stasiak,~A. \emph{Macromolecules} \textbf{2008}, \emph{41}, 4444\relax
\mciteBstWouldAddEndPuncttrue
\mciteSetBstMidEndSepPunct{\mcitedefaultmidpunct}
{\mcitedefaultendpunct}{\mcitedefaultseppunct}\relax
\EndOfBibitem
\bibitem[Livingston(1993)]{livingston1993knot}
Livingston,~C. \emph{Knot Theory}; Carus Monographs; Mathematical Association
  of America, 1993; Vol.~24\relax
\mciteBstWouldAddEndPuncttrue
\mciteSetBstMidEndSepPunct{\mcitedefaultmidpunct}
{\mcitedefaultendpunct}{\mcitedefaultseppunct}\relax
\EndOfBibitem
\bibitem[Gutjahr et~al.(2006)Gutjahr, Lipowsky, and
  Kierfeld]{lipowsky:epl:2006}
Gutjahr,~P.; Lipowsky,~R.; Kierfeld,~J. \emph{Europhys. Lett.} \textbf{2006},
  \emph{76}, 994\relax
\mciteBstWouldAddEndPuncttrue
\mciteSetBstMidEndSepPunct{\mcitedefaultmidpunct}
{\mcitedefaultendpunct}{\mcitedefaultseppunct}\relax
\EndOfBibitem
\bibitem[Arai et~al.(1999)Arai, Yasuda, Akashi, Harada, Miyata, {Kinoshita,
  Jr.}, and Itoh]{arai:nature:1999}
Arai,~Y.; Yasuda,~R.; Akashi,~K.; Harada,~Y.; Miyata,~H.; {Kinoshita, Jr.},~K.;
  Itoh,~H. \emph{Nature} \textbf{1999}, \emph{399}, 446\relax
\mciteBstWouldAddEndPuncttrue
\mciteSetBstMidEndSepPunct{\mcitedefaultmidpunct}
{\mcitedefaultendpunct}{\mcitedefaultseppunct}\relax
\EndOfBibitem
\bibitem[Bao et~al.(2003)Bao, Lee, and Quake]{bao:prl:2003}
Bao,~X.~R.; Lee,~H.~J.; Quake,~S.~R. \emph{Phys. Rev. Lett.} \textbf{2003},
  \emph{91}, 265506\relax
\mciteBstWouldAddEndPuncttrue
\mciteSetBstMidEndSepPunct{\mcitedefaultmidpunct}
{\mcitedefaultendpunct}{\mcitedefaultseppunct}\relax
\EndOfBibitem
\end{mcitethebibliography}

\end{document}